\documentclass[showpacs,pre,aps,twocolumn,superscriptaddress]{revtex4-1}

\pdfoutput=1

\usepackage{amsmath,amsfonts,amssymb,bm,graphicx,hyperref}

\begin{document}

\newcommand{\beq}{\begin{equation}}
\newcommand{\eeq}{\end{equation}}

\title{Cycle counts and affinities in stochastic models of non-equilibrium systems}

\author{Patrick Pietzonka}
\affiliation{Department of Applied Mathematics and Theoretical Physics, University of Cambridge, Wilberforce Road, Cambridge CB3 0WA, United Kingdom}

\author{Jules Guioth}
\affiliation{Department of Applied Mathematics and Theoretical Physics, University of Cambridge, Wilberforce Road, Cambridge CB3 0WA, United Kingdom}
\affiliation{Univ. Lyon, \'{E}NS de Lyon, Univ. Claude Bernard, CNRS, Laboratoire de Physique, F-69342 Lyon, France}

\author{Robert L. Jack}
\affiliation{Department of Applied Mathematics and Theoretical Physics, University of Cambridge, Wilberforce Road, Cambridge CB3 0WA, United Kingdom}
\affiliation{Yusuf Hamied Department of Chemistry, University of Cambridge, Lensfield Road, Cambridge CB2 1EW, United Kingdom}

\begin{abstract} 
For non-equilibrium systems described by finite Markov processes, we consider the number of times that a system traverses a cyclic sequence of states (a cycle).  The joint distribution of the number of forward and backward instances of any given cycle is described by universal formulae which depend on the cycle affinity, but are otherwise independent of system details.  We discuss the similarities and differences of this result to fluctuation theorems, and generalize the result to families of cycles, relevant under coarse-graining.  Finally, we describe the application of large deviation theory to this cycle counting problem.
\end{abstract}

\maketitle

\section{Introduction}
\label{sec:intro}

Fluctuations in non-equilibrium systems continue to provide surprises and new insight in statistical physics.  
Among the most famous examples are fluctuation theorems~\cite{Gallavotti1995,Jarzynski1997,Crooks2000,AG,Seifert2012}, which come in different types.  Some of them 
(detailed fluctuation theorems) are symmetries of probability distributions~\cite{Gallavotti1995,Crooks2000,AG}, while others allow the results of dynamical experiments to be related to (static) quantities such as free energies~\cite{Jarzynski1997,Crooks2000}.  More recently, thermodynamic uncertainty relations (TURs) have been derived~\cite{bara15,ging16,piet15,horo19,Dechant2020}, which are inequalities that relate the variances of physical observables to underlying properties of a system, particularly its entropy production.

These results reflect an elegant mathematical structure, that underpins the physical models to which they apply.  They also have experimental relevance~\cite{Liphardt2002,cili17}.  Still, it is notable that fluctuation theorems for non-equilibrium steady states usually involve quantities such as affinities or the entropy production, which are difficult to characterise from experimental (or simulation) data.  There is an ongoing effort to infer such quantities by TURs~\cite{bara15a,li19,mani20}.

In this context, a recent result of Biddle and Gunawardena (BG)~\cite{BG} offers a potentially new route towards inference of cycle affinities from data.   Analyses of steady-state fluctuations often focus on currents, but BG's results concern cycles -- sequences of states that begin and end at the same point.  They showed that for long times, the affinity of a cycle can be computed by counting the number of times that the cycle is traversed, in each direction. 
In their approach,
completion of a cycle corresponds to the system visiting a particular set of states, in a particular order.   
This distinguishes their analysis from a different body of work, that involves decomposition of the full stochastic trajectory into a sequence of completed cycles~\cite{qian-book,kalpazidou-book,Jia2016}: in that case, a system's progress around a single cycle may include other cycles within it, see also~\cite{polettini2021} for a recent example of this decomposition in practice.   
Yet another approach~\cite{AG} is based on Schnakenberg network theory~\cite{schn76}, {where the probability currents between pairs of states are decomposed into a minimal set of cycle currents}, see also~\cite{Altaner2012,Bertini2015-spa}.  However, that construction does not provide information about {any specific} cyclic sequence of states.

Given their definition of a cycle as a specific sequence of states, the approach of~\cite{BG} can be interpreted as a generalization to \emph{continuous-time} Markov chains of a set of discrete-time problems, including counting the number of occurrences of a given word, in a random sequence of letters.  Such problems are relevant for  
DNA sequence analysis~\cite{Schbath1997,robin_daudin_1999,Lothaire2005,Roquain2007}.  
There are also physical results for cycle counting in discrete-time, at least for unicyclic models~\cite{Roldan2019}.

In this paper, we follow~\cite{BG}, exploring in more detail the distribution of the number of times that cycles are traversed, in either direction.
The results apply at all times.  Hence they generalize the results of~\cite{BG}, which concern the mean of the distribution, for long trajectories.  We also outline methods for analysis of large-deviation events~\cite{Touchette2009,Chetrite2015,Jack2020}, where the cycle count takes a non-typical value at very large times.   As might be expected, the cycle affinity plays a central role in the statistics of cycle counts, especially for the cycle current, which is the difference in counts for forward and reverse cycles.  By contrast, the statistics of the total count have a complex dependence on all model parameters, as expected for time-reversal symmetric (frenetic) quantities~\cite{Maes2020}.   
Some of the methods used here, particularly renewal theory and $m$th order Markov processes, are similar to those used for sequence analysis, although our results concern processes in continuous time.

The form of the paper is as follows: Sec.~\ref{sec:def} defines the models and quantities of interest and Sec.~\ref{sec:results} describes the main results for fluctuations of cycle counts, in finite time intervals.  Sec.~\ref{sec:discuss-finite} discusses these results, including some possible extensions, and the connection with fluctuation theorems.  The relevant theory for large deviations of cycle counts is outlined in Sec.~\ref{sec:large}.
Finally, Sec.~\ref{sec:conc} gives a short conclusion.

\section{Model and Definitions}
\label{sec:def}

\newcommand{\as}{\mathrm{A}}
\newcommand{\bs}{\mathrm{B}}
\newcommand{\cs}{\mathrm{C}}
\newcommand{\ds}{\mathrm{D}}

\newcommand{\CC}{\mathcal{C}}
\newcommand{\CR}{\mathcal{C}^{\rm R}}

\newcommand{\Ac}{\mathcal{A}}

\newcommand{\nhat}{\hat{n}}
\newcommand{\nR}{n^{\rm R}}
\newcommand{\nRhat}{\hat{n}^{\rm R}}

\newcommand{\khat}{\hat{k}}
\newcommand{\jhat}{\hat{\jmath}}

\newcommand{\tobs}{\tau}

\begin{figure}
\includegraphics[width=8.5cm]{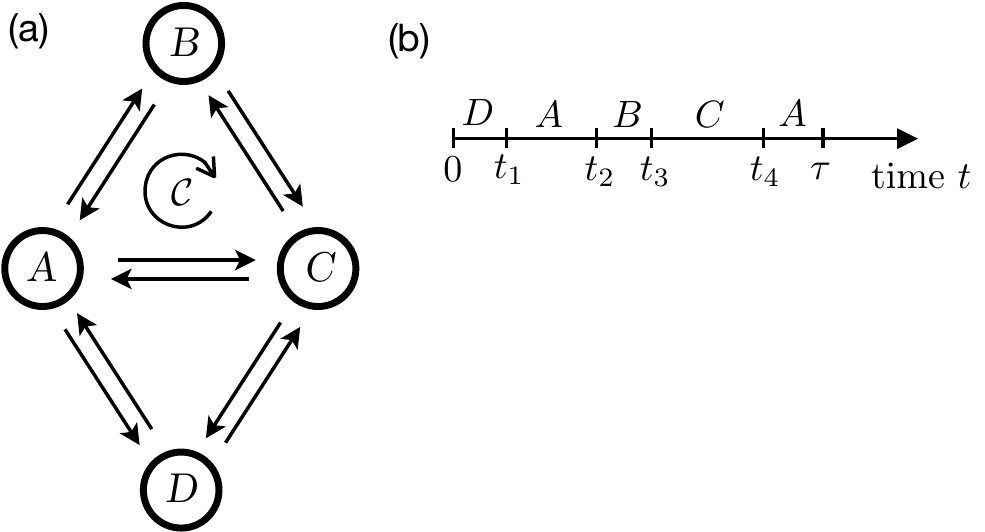}
\caption{(a) Diagram showing a simple system of four states, with transitions indicated by straight arrows.  The cycle $\CC=\as\bs\cs\as$ is also indicated. (b) Example trajectory for a time period $[0,\tau]$, with jumps between states at times $
\bm{t}_X=(t_1,t_2,t_3,t_4)$.  The sequence of states is $\bm{x}_X = (\ds,\as,\bs,\cs,\as)$ so $n_\CC(X)=1$ and $n_\CR(X)=0$.}
\label{fig:sketch1}
\end{figure}

\emph{Model}
--
We consider non-equilibrium systems that are modelled as Markov chains with discrete states, in continuous time.  
A very simple example system is shown in Fig.~\ref{fig:sketch1}(a).  The set of states is denoted by $\Gamma$, which in the example is $\{\as,\bs,\cs,\ds\}$.  
The transition rate from state $x$ to state $y$ is denoted by $w(x\to y)\geq 0$. 
We emphasise that  the results apply for any finite set $\Gamma$ so they include complex systems like (finite) exclusion processes and are not at all restricted to simple models like the example of Fig.~\ref{fig:sketch1}.  

We restrict to irreducible systems and we also assume that if $w(x\to y)>0$ then also $w(y\to x)>0$ (which is sometimes called weak reversibility or microreversibility).  This ensures that the steady state is unique and that every state $x$ has a non-zero probability in the steady state, denoted by $\pi(x)$.  Also the average entropy production rate is finite and non-negative.

\emph{Trajectories}
--
A trajectory of the system can be specified for a time period $[0,\tau]$, an example is shown in Fig.~\ref{fig:sketch1}(b).   Let $X$ denote a trajectory, it
consists of a sequence of states 
\beq
\bm{x}_X=(x_0,x_1,x_2,\dots,x_M)
\eeq 
and the associated transition times 
\beq
\bm{t}_X=(t_1,t_2,\dots,t_{M})
\eeq
with $0<t_1<t_2<\cdots<t_{M}<\tobs$.  Here each $x_i\in\Gamma$, the system starts in state $x_0$ and jumps to state $x_i$ at time $t_i$.  The number of transitions $M$ is a random (trajectory-dependent) quantity.

\emph{Words and cycles}
--
We consider sequences of states visited by the Markov chain.
Similar objects have been studied for models in discrete time, especially in the context of DNA sequence analysis~\cite{Schbath1997,robin_daudin_1999,Lothaire2005,Roquain2007}.  The continuous-time case is very similar, although it requires some additional book-keeping.

A sequence of states is called a word.  We always restrict to words that can occur in trajectories of the system. (In the example of Fig.~\ref{fig:sketch1}, $\as\bs$ or $\bs\as\cs$ would be suitable words, but $\as\bs\ds$ is excluded because $w({\bs\to\ds})=0$.)  
A word that begins and ends with the same letter is called \emph{a cycle}, and if $\CC$ is a cycle then $\CR$ denotes its time-reversal.  For example if $\CC=\as\bs\cs\as$ then $\CR=\as\cs\bs\as$.   The cycle length $m_\CC$ is the number of transitions required to complete a cycle, so $m_\CC=3$ in this example (one fewer than the corresponding word length).  

Denote by $\CC_j$ the $j$th state in cycle $\CC$, so $1\leq j \leq m_\CC+1$. Then the cycle affinity for $\CC$ is
\beq
\Ac_\CC = \sum_{j=1}^{m_\CC} \ln \frac{ w(\CC_j\to \CC_{j+1}) }{ w(\CC_{j+1}\to \CC_{j})  },
\label{equ:aff}
\eeq
which is also the entropy production for one cycle in the steady state.  For allowed cycles then all rates are non-zero (by the weak reversibility property) so the affinity is finite.

In a trajectory, it is convenient to define the \emph{start time} of a word (or cycle) as the time of the jump between the first two states, and the \emph{end time} as the time of the jump between the last two states.  So for $\as\bs\cs\as$ the word starts at the transition $\as\to\bs$ and ends at $\cs\to\as$.  The \emph{completion time} is the difference between the start and end time.

{As noted in Sec.~\ref{sec:intro}, this definition of a cycle~\cite{BG} (which might also be called a \emph{cyclic word}) -- and the corresponding start/end times -- differ from the definitions used in other works such as~\cite{Ge2008,Ge2012,Jia2016,polettini2021}.  Our definition leads to a simpler analysis, but some of the results and methods are similar.}

\emph{Counting cycles}
--
Given a trajectory $X$ and a cycle $\CC$, write $\nhat_\CC(X)$ for the number of occurrences of $\CC$ in $X$.   This is the number of times that the word $\CC$ appears in the sequence $\bm{x}_X$.  (The hat serves as a reminder that $\hat{n}_\CC$ is a random variable.)  The cycle must appear exactly as in its definition, and different occurrences of the cycle may overlap.  (For example the cycle $\as\bs\as\bs\as$ appears twice in the sequence $\as\bs\as\bs\as\bs\as$.)  Of course, generic words can be counted in the same way, not only cycles.  It is convenient to write $\nRhat_\CC = \nhat_{{\CR}}$ for the number of occurrences of the reverse cycle.

\emph{Non-revisiting cycles}
--
It will be convenient in the following to distinguish two kinds of cycle.  Recalling that a cycle always begins and ends at the same point, we define a non-revisiting cycle as one in that does not return to its initial point, until the end.  For example $\as\bs\as$ is a non-revisiting cycle but $\as\bs\cs\as\cs\bs\as$ is not.  One sees that different occurrences of a non-revisiting cycle cannot overlap each other, and that a general cycle can be decomposed as the concatenation of non-revisiting cycles.  The class of non-revisiting cycles is larger than that of simple cycles (for example $\as\bs\cs\bs\cs\as$ is non-revisiting), but it is not as large as the class of non-overlapping cycles (or words) from~\cite{robin_daudin_1999}.

\section{results: finite time}
\label{sec:results}

This Section contains some general results for the probability distribution of the number of cycle counts, for finite trajectories with time $t\in[0,\tau]$.  We first summarise the results before giving the derivations.  The analysis leading to these results is quite straightforward, but we argue that the results are interesting for two reasons: first, as a possible way to infer model parameters (specifically, affinities) from data~\cite{BG}; and also as a starting point for more detailed analysis of cycle counts.  Both these directions are discussed in later Sections.

\subsection{Overview}
\label{sec:overview}

Our results concern the random variables $\nhat_\CC,\nRhat_\CC$, for cycles as defined above.  
In the palindromic case $\CC=\CR$ then none of these results have any content, so we assume throughout that $\CC\neq\CR$.
We are motivated by a result of~\cite{BG}, which is that for long trajectories
\beq
\lim_{\tobs\to\infty} \frac{ \nhat_\CC(X) }{ \nRhat_\CC(X) } = {\rm e}^{ \Ac_\CC } \; .
\label{equ:BG}
\eeq
Such formulae require some care because the left hand side is a random (trajectory-dependent) quantity but the right hand side is deterministic: the equation holds in the same sense as a law of large numbers.  
The physical idea behind (\ref{equ:BG}) is that the cycle affinity can be inferred by counting cycles that are traversed in forward and backward directions.  

In the following, we derive several results, related to (\ref{equ:BG}).  
%We summarize them here, before proceeding with the analysis.
First, the derivation of~\cite{BG} can be easily generalized to obtain a result for steady-state averages over trajectories of finite length $\tobs$, {with arbitrary initial condition}.  The result is
\beq
\frac{\left\langle \nhat_\CC  \right\rangle }{\left\langle \nRhat_\CC \right\rangle} = {\rm e}^{ \Ac_\CC } 
\label{equ:BG-ave}
\eeq
where $\langle\cdot\rangle$ indicates an average over trajectories of the system (the dependence of the cycle counts on $X$ is implicit).  This result states that cycles with positive affinity happen more often in the forward direction, as expected.  As $\tau\to\infty$ then $\nhat_\CC(X)\to\left\langle \nhat_\CC  \right\rangle$ with probability one, this is a weak law of large numbers.   The proposal of~\cite{AG} was that \eqref{equ:BG} might be used to infer affinities from data; in this case \eqref{equ:BG-ave} seems also useful since long trajectories are not required.   

We note in passing that the mean number of cycles is not a simple linear function of the trajectory length, that is $\left\langle \nhat_\CC  \right\rangle \neq \tau \omega(\CC)$ in general. (Here $\omega(\CC)$ would be a cycle completion rate.)   The reason is that there is typically a significant lag time between starting and ending a cycle.  So the fact that (\ref{equ:BG-ave}) applies for all $\tau$ is not trivial.

We now consider the joint distribution of $n_\CC,\nR_\CC$, which we denote by
$P_\tobs(n_\CC,\nR_\CC)$.  
Our results for this distribution are restricted to non-revisiting cycles, but they hold for any trajectory length $\tau$
and for any initial condition (it is not necessary that the probabilities are evaluated in the steady state of the system).
We show that
\beq
n_\CC P_\tobs(n_\CC,\nR_\CC-1) =  {\rm e}^{ \Ac_\CC } \nR_\CC P_\tobs(n_\CC-1,\nR_\CC) .
\label{equ:FT-simple}
\eeq
The physical origin of \eqref{equ:FT-simple} is that replacing any non-revisiting cycle $\CC$ by its time-reversed counterpart $\CR$ changes the trajectory probability by a factor ${\rm e}^{-\Ac_\CC}$. The prefactors $n_\CC$ and $\nR_\CC$ are of combinatorial origin.

Also, it is straightforward to show that for non-revisiting cycles
\beq
P_\tobs(n_\CC,\nR_\CC) 
=  P_\tobs(\nR_\CC, n_\CC) \exp\left[ (n_\CC-\nR_\CC) {\cal A}_\CC \right]  \; ,
\label{equ:FT-all}
\eeq
which has some similarities to the fluctuation theorem of Andrieux and Gaspard~\cite{AG}, see Sec.~\ref{sec:fluct}.

Both \eqref{equ:FT-simple} and \eqref{equ:FT-all} are direct consequences of a binomial structure in the distribution $P_\tobs(n_\CC,\nR_\CC)$. In order to state this property conveniently, identify the total number of cycles in trajectory $X$ and the corresponding net flux as
\begin{align}
\hat{K}_\CC(X) & = { \nhat_\CC(X)  + \nRhat_\CC(X)  }
\nonumber\\
\hat{J}_\CC(X) & = { \nhat_\CC(X)  - \nRhat_\CC(X)  }.
\label{equ:JK}
\end{align}
Denoting the joint distribution of these quantities by $\tilde{P}_\tau(K,J)$ {and the marginal of $K$ by $\tilde{P}_{\tau}(K)= \sum_J \tilde P_\tau(K,J)$}, we show in Sec.~\ref{sec:proof} that the conditional distribution of $\hat J$, {namely $\tilde{P}_\tau(K,J)/\tilde{P}_\tau(K) $}, is binomial, so that
\beq
\tilde{P}_\tau(K,J) = \tilde{P}_\tau(K) \begin{pmatrix}  K \\ \frac12(K+J) \end{pmatrix} 
\frac{\exp({J\Ac/2}) } {  [2\cosh(\Ac/2)]^K }  \; .
\label{equ:P-KJ-bin}
\eeq
Here and in the following, we sometimes omit the label $\CC$ for variables and affinities, where there is no ambiguity.
The key point of~\eqref{equ:P-KJ-bin} is that the dependence of $\tilde{P}_\tau$ on $J$ is explicit. This formula holds for all models and for any non-revisiting cycle $\CC$.  In discrete time, a similar result is given in~\cite{Roldan2019}, for the restricted case of unicyclic models.

These results extend the analysis of~\cite{BG} from the most likely number of cycles to its full fluctuation spectrum.  However, they are restricted to non-revisiting cycles: {we emphasise that (\ref{equ:FT-simple},\ref{equ:FT-all}) are derived from the more general (\ref{equ:P-KJ-bin}), so this restriction is necessary for all these results}.
For such cycles, one may then recover previous results for the mean, in particular (\ref{equ:BG-ave}) is obtained by summing both sides of \eqref{equ:FT-simple} over $n_\CC,\nR_\CC$.

It is useful to recall that the path weight of trajectory $X$ in our general model is 
\begin{align}
P[X]=P_0(x_0)&\left[\prod_{i=0}^{M-1} w(x_i\to x_{i+1}){\rm e}^{-r(x_i) (t_{i+1}-t_{i})}\right]\nonumber\\&\qquad\times{\rm e}^{-r(X_M) (\tau-t_{M})}
\label{eq:pathweight}
\end{align}
{where $t_0=0$}; also $P_0(x_0)$ is the (arbitrary) distribution of the initial state and we introduced the exit rate from state $x$, as % $r(x)$ defined in~\eqref{equ:exit-rate} and 
\beq
r(x)=\sum_{y(\neq x)}w(x\to y) \; .
\label{equ:exit-rate}
\eeq
From this formula, the result (\ref{equ:FT-simple}) can be anticipated by observing that any instance of a non-revisiting cycle $\CC$ in $X$ can be replaced by an instance of $\CR$, so that $P[X]$ changes by a factor ${\rm e}^{-\cal A_\CC}$.  A precise argument along these lines is given in Sec.~\ref{sec:proof}, see also Fig.~\ref{fig:sketch2}.  {Alternatively, this result [and also (\ref{equ:P-KJ-bin})] may be derived using renewal theory, see Appendix~\ref{app:renew}.}

\subsection{Average cycle counts}

We now derive (\ref{equ:BG-ave}).
Consider the probability that an instance of cycle $\CC$ starts between times $t$ and $t+\delta t$ and ends before time $\tau$.  For small $\delta t$ denote this by $P_\CC(t,\tau) \delta t$.  Considering trajectories for the time period $[0,\tau]$, the average
 $\langle \hat{n}_\CC \rangle$ may then be decomposed as
\beq
\langle \hat{n}_\CC \rangle = \int_0^\tau P_\CC(t,\tau) dt \; .
\label{equ:ave-lin}
\eeq
Moreover, the probability that a transition $\CC_1 \to \CC_2$ occurs between times $t$ and $t+\delta t$ is $p(\CC_1,t) w(\CC_1 \to \CC_2) \delta t$, {where $p(\CC_1,t)$ is the probability that the system is in state $\CC_1$ at time $t$}.  Since every instance of cycle $\CC$ starts with such a transition, it follows that
\beq
P_\CC(t,\tau) = { p(\CC_1,t) } w(\CC_1 \to \CC_2) p_{\rm seq}(\CC) F_{\rm fin}(\tau-t,\CC),
\label{equ:PC}
\eeq
where
\beq
p_{\rm seq}(\CC) = \prod_{j=2}^{m_\CC} \frac{w(\CC_j\to \CC_{j+1})}{r(\CC_j)}
\eeq
is the probability that the system follows the correct sequence of states, and
the factor $F_{\rm fin}(\tau-t,\CC)$ in (\ref{equ:PC})
is the probability that the cycle is completed in a time less than $\tau-t$.  [This cycle completion time is a sum of $m_\CC-1$ exponentially distributed random variables with means $r(\CC_2)^{-1},r(\CC_3)^{-1},\dots,r(\CC_{m_{\CC}})^{-1}$.]  
Hence
\begin{multline}
\langle \hat{n}_\CC \rangle =  \int_0^\tau \Big[ p(\CC_1,t) r(\CC_1)  \prod_{j=1}^{m_\CC} \frac{w(\CC_j\to \CC_{j+1})}{r(\CC_j)} 
\\ \times   F_{\rm fin}(\tau-t,\CC) \Big]  dt \; .
\label{equ:mean-n}
\end{multline}

Repeating the same argument for the reversed cycle $\CR$ one finds 
\begin{multline}
\langle \nRhat_\CC \rangle = \int_0^\tau \Big[  p(\CC_1,t) {r(\CC_1)} \prod_{j=1}^{m_\CC} \frac{w(\CC_{j+1}\to \CC_{j})}{r(\CC_{j+1})}
 \\ \times F_{\rm fin}(\tau-t,\CR)  \Big]  dt \; .
 \label{equ:mean-nR}
\end{multline}
Note that the time to complete cycle $\CR$ is the same sum of exponentially distributed random variables as for $\CC$, so 
\beq
F_{\rm fin}(\tau-t,\CR) = F_{\rm fin}(\tau-t,\CC) \; ,
\label{equ:F-fin-R}
\eeq 
{see also~\cite{Ge2008,Ge2012}}, {an explicit formula for $F_{\rm fin}$ is given in \eqref{equ:F-fin-explicit}}.
Also, the fact that the cycle starts and ends at the same point means that $\prod_{j=1}^{m_\CC} r(\CC_{j+1}) = \prod_{j=1}^{m_\CC} r(\CC_{j}) $.
Combining these facts with (\ref{equ:aff},\ref{equ:mean-n},\ref{equ:mean-nR}), one recovers (\ref{equ:BG-ave}).

Note that there is no restriction here to non-revisiting cycles, the physical reason is that (\ref{equ:ave-lin}) decomposes the mean number of cycles into a sum of independent averages. {To see this, recall that the average number of cycles that start between time $t$ and $t+\delta t$ and end before time $\tau$ is $P_\CC(t,\tau)\delta t$.  Integrating over $t$ corresponds to summing these independent averages and gives the average number of completed cycles within the full trajectory.}  The possibility of overlapping cycles is important for fluctuations in their number, but not for the mean.

\subsection{Fluctuations for non-revisiting cycles}
\label{sec:proof}

\begin{figure}
\includegraphics[width=8.5cm]{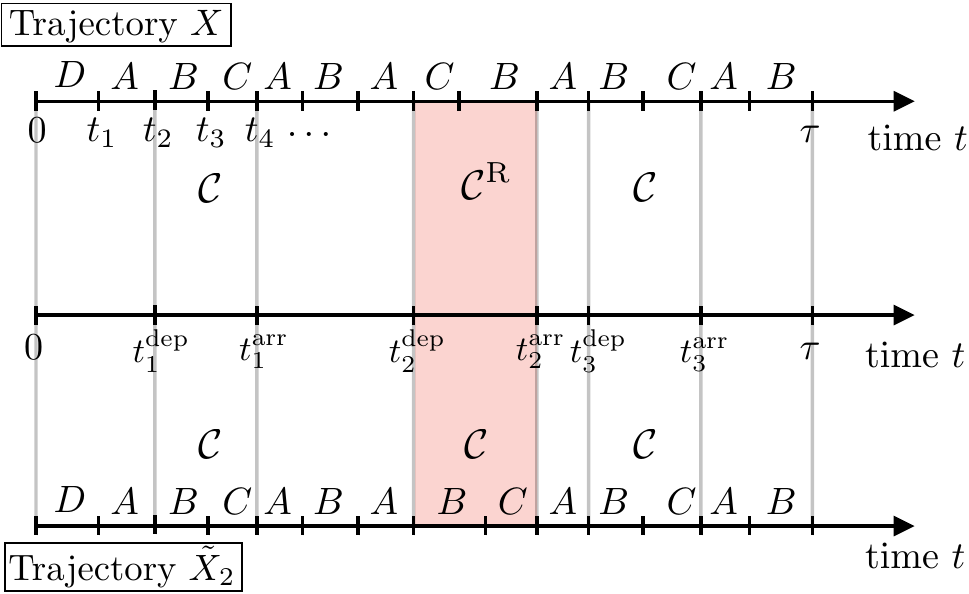}
\caption{(Top) A trajectory $X$ of the four-state system in Fig.~\ref{fig:sketch1}, and the corresponding sequence of completions of the cycle $\CC$ or $\CR$.
We keep track of the departure and arrival times for state $A$ that enclose instances of $\CC$ or $\CR$.  The second such instance is highlighted in red. (Bottom) Applying time-reversal to the highlighted part of the trajectory leads to the trajectory $\tilde X_2$, transforming the cycle $\CR$ into $\CC$.}
\label{fig:sketch2}
\end{figure}

We now restrict to non-revisiting cycles, and we derive (\ref{equ:FT-simple}-\ref{equ:P-KJ-bin}).  We use a methodology similar to proofs of fluctuation theorems based on path weights~\cite{Seifert2012}, see Appendix~\ref{app:alt_proof} for a derivation using concepts of renewal theory.  

Suppose that  the cycle $\CC$ of interest starts in state $\as$ (this does not lose any generality).  For any trajectory $X$ we can identify the sequence of completions of the cycle in either forward or backward direction, for example
\beq
{\cal S}[X]=(\CC,\CR,\CC,\CC,\CR),
\label{equ:seq-example}
\eeq
along with the sequence of start and end times of the cycles, denoted by $(t^{\rm dep}_1,t^{\rm dep}_2,\dots)$ and $(t^{\rm arr}_1,t^{\rm arr}_2,\dots)$ respectively, see Fig.~\ref{fig:sketch2}.  (The start/end times of the cycle correspond to the departure/arrival times from/to state $\as$.) The probability to observe a specific sequence ${\cal S}$ of forward and backward cycles within time $\tau$ can be expressed as
\beq
{\cal P}_\tau({\cal S})=\sum_X P[X] \delta({\cal S},{\cal S}[X]),
\label{eq:seqprob}
\eeq
where the sum runs over all trajectories of length $\tau$, suitably parameterised as a path integral. The function $\delta({\cal S},{\cal S}')=1$ if ${\cal S}={\cal S}'$, and zero otherwise.

To derive (\ref{equ:P-KJ-bin}), we first obtain formulae that relate the probabilities of specific trajectories; then we sum over (classes of) trajectories to obtain the distribution of $(K,J)$.  
Note that the sequence ${\cal S}[X]$ consists of $\hat K(X)$ elements.
For every trajectory $X$ we define a conjugate trajectory $\tilde X_k$ as follows: If $k\leq \hat K(X)$ then $\tilde{X}_k$ is obtained from $X$ by reversing {in time} the $k$th cycle in ${\cal S}[X]$, that is:
\beq
\tilde{X}_k(t)=\begin{cases}X(t^{\rm arr}_k-t+t^{\rm dep}_k),&\textup{if }t^{\rm dep}_k<t<t^{\rm arr}_k \\X(t),&\textup{otherwise}\end{cases}.
\label{eq:partial_timerev}
\eeq
{%
See Fig.~\ref{fig:sketch2}, {which shows how $\tilde X_{k}$ is obtained from $X$ for $k=2$ by reversing the second instance of the cycle}.
Similar partial time-reversal operations have been considered before in the context of simple chemical reactions~\cite{Ge2008,Ge2012} as well as in other works on cycle counting~\cite{Jia2016,polettini2021}.
It is convenient to extend this definition to include $k>\hat K(X)$ by taking $\tilde{X}_k(t)=X(t)$.
}

Noting that the sojourn times in each state are unaffected by the partial time-reversal, we see from \eqref{eq:pathweight} and \eqref{equ:aff} that
\beq
\frac{P[X]}{P[\tilde X_k]}=\begin{cases} {\rm e}^{\pm{\cal A_C}}, &\textup{if } k\leq \hat K(X) 
\\ 1 &\textup{otherwise} \end{cases},
\label{equ:detail-m}
\eeq
where in the first case we take the plus sign if the $k$th entry in ${\cal S}(X)$ is ${\cal C}$, and the minus sign if this entry is $\CR$.  

  Now define $\tilde{\cal S}_k[X] = {\cal S}[\tilde X_k]$; for example if ${\cal S}=(\CC,\CR,\CC,\CC,\CR)$ and $k=2$ then $\tilde{\cal S}_2=(\CC,\CC,\CC,\CC,\CR)$.
The mapping between $X$ and $\tilde X_k$ is a bijection, which means that (\ref{eq:seqprob}) can be expressed as 
\begin{align}
{\cal P}_\tau({\cal S}) & =\sum_X P[\tilde X_k] \delta({\cal S},{\cal S}[\tilde X_k]) 
\nonumber\\
& =\sum_X P[\tilde X_k] \delta({\cal S},\tilde{\cal S}_k[X]),
\end{align}
where the first equality is obtained by relabelling the trajectories and the second uses the definition of $\tilde{\cal S}$.  Using (\ref{equ:detail-m}) yields 
\beq
{\cal P}_\tau({\cal S})={\rm e}^{\pm{\cal A_C}}{\cal P}_\tau(\tilde{\cal S}_k),
\label{eq:detailed_FT}
\eeq
where we take the plus (or minus) sign if the $k$th entry in ${\cal S}(X)$ is ${\cal C}$ (or $\CR)$, as in (\ref{equ:detail-m}). (It is assumed that the number of entries in ${\cal S}$ is at least as large as $k$.)
The result (\ref{eq:detailed_FT}) is a special example of a detailed fluctuation theorem~\cite{Seifert2012}.

Now, given a sequence ${\cal S}$, one may use (\ref{eq:detailed_FT}) and successively replace all instances of $\CR$ by $\CC$ to obtain
\beq
{\cal P}_\tau({\cal S})={\rm e}^{-\nR_\CC {\cal A_\CC}}{\cal P}_\tau(\CC,\CC,\dots,\CC)
\label{equ:detail-S}
\eeq
where  $\nR_\CC$ is the number of occurrences of $\CR$ in the sequence ${\cal S}$.  Write $K_{\cal S}$ for the number of entries in ${\cal S}$.   Then
the right hand side of (\ref{equ:detail-S}) is the probability of $K_{\cal S}$ forward cycles and no reverse ones, that is $P_\tau(K_{\cal S},0)$ in the notation of (\ref{equ:FT-simple}). The probability $P_\tau(K_{\cal S}-n,n)$ can be obtained by summing (\ref{equ:detail-S}) over sequences ${\cal S}$ with the requisite numbers of forward and reverse cycles: the number of elements in the sum is a binomial coefficient.  One obtains
\begin{align}
P_\tau(n_\CC,\nR_\CC)%&=\sum_{{\cal S}|n_\CC,\nR_\CC}{\cal P}_\tau({\cal S})\nonumber\\&
=
\begin{pmatrix}
   n_\CC+\nR_\CC\\n_\CC
\end{pmatrix}
{\rm e}^{-\nR_\CC {\cal A_C}}P_\tau(n_\CC+\nR_\CC,0).
\end{align}
Re-parameterisation in terms of $K,J$ yields~\eqref{equ:P-KJ-bin}. 
As noted in Sec.~\ref{sec:overview}, both~\eqref{equ:FT-simple} and~\eqref{equ:FT-all} follow straightforwardly from this result.

Finally, we observe that while these results have been derived for non-revisiting cycles, this is not the most general case in which (\ref{eq:detailed_FT}) and hence (\ref{equ:P-KJ-bin}) apply.  Eq.~(\ref{eq:detailed_FT}) does not apply for \emph{all} cycles because if two instances of the same cycle can overlap each other, then it is not generally possible to reverse a single instance of the cycle, leaving all other instances unchanged.  [For example, consider the (revisiting) cycle ABCABCA and a trajectory that contains the sequence ABCABCABCA.]  A similar problem arises if an instance of $\CC$ can overlap with an instance of $\CR$.  The assumption of non-revisiting cycles is sufficient to ensure that such overlaps never occur and (\ref{eq:detailed_FT}) holds, but this condition is not necessary.  Classes of non-overlapping words are discussed (for example) in~\cite{robin_daudin_1999}; we avoid such issues here, to simplify the analysis.

\section{Discussion of finite-time results}
\label{sec:discuss-finite}

\subsection{Coarse-grained measurements: families of cycles}
\label{sec:generalisation}

The results derived so far concern the statistics of completions of a given cycle $\CC$, which is a specific sequence of states.   Hence, any measurement of $n_\CC$ requires complete information about the trajectory of the system.  Since we consider ``mesoscopic'' models that should be defined as coarse-grained representations of real physical systems, it is useful to consider how this requirement of complete information can be reconciled with a coarse-graining operation.

Note first that these results can be generalised to some situations where incomplete information is available.  To see this, let  $\cal F$ represent a family (a set) of cycles, and let 
\beq
\hat{n}_{\cal F}(X) = \sum_{\CC \in {\cal F}} \hat{n}_\CC(X)
\eeq
be the total number of occurrences in trajectory $X$ of all cycles $\CC$ from that family $\cal F$. {Reversing all cycles in $\cal F$ yields the family ${\cal F}^{\rm R}$, with the number of occurrences $\hat{n}_{\cal F}^{\rm R}$ defined analogously.  {(We assume that if $\CC\in\cal F$ then $\CR\notin\cal F$.)}}

If all members of $\cal F$ have the same affinity ${\cal A}_{\cal F}$ then it is obvious that (\ref{equ:BG-ave}) still holds (with $\hat{n}_{\cal C}, \nRhat_\CC,{\cal A}_\CC$ replaced by $\hat{n}_{\cal F}, \nRhat_{\cal F},{\cal A}_{\cal F}$).  If all members of $\cal F$ are also non-revisiting then (\ref{equ:FT-simple}-\ref{equ:P-KJ-bin}) hold too.  [This can be seen by constructing a modified sequence ${\cal S}[X]$ in which the symbol ${\cal C}$ represents a completion of any member of ${\cal A}_{\cal F}$ and $\CR$ represents
completion of the any member of the family ${\cal F}^{\rm R}$.   Then (\ref{eq:detailed_FT}) holds and the analysis follows.]

A simple example of such a family is obtained by including repeated forwards and backwards steps within the cycle.  For example, consider the family containing
$\as\bs\cs\as ,  \as\bs\cs\bs\cs\as$, and all similar cycles obtained by repeatedly inserting instances of $\bs\cs$ before the final $\as$.  All these cycles obviously have the same affinity and they are non-revisiting, so (\ref{equ:BG-ave}-\ref{equ:P-KJ-bin}) still hold for the joint distribution of $(\hat{n}_{\cal F},\hat{n}_{\cal F}^{\rm R})$.  
{[To connect the results here with the framework of~\cite{kalpazidou-book,Jia2016}, it is necessary to consider larger families, which include cycles that are constructed from a main (outer) cycle, and also include non-trivial subcycles; one should also extend the definition of a time-reversed cycle appropriately, so that only the main cycle is reversed, leaving the subcycles invariant.  Such complex families are not our main concern in this work.]}

\begin{figure}
\includegraphics[width=3.5cm]{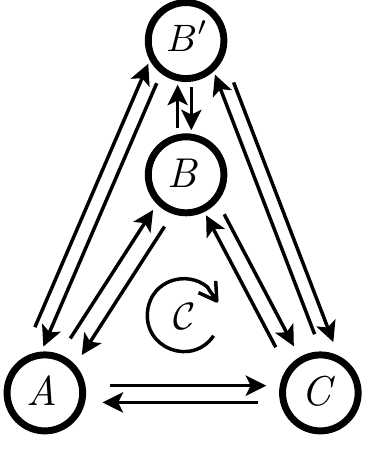}
\caption{A network where $B$ and $B'$ form a pair of sub-states. If the physical driving mechanism does not distinguish between these sub-states, variations of the cycle $\cal C$ visiting either $B$ or $B'$ all have the same affinity and can be lumped together in a single family of cycles.}
\label{fig:cycle_families}
\end{figure}

Families of cycles with equal affinity also arise naturally in physical situations, especially where coarse-graining is considered. 
%This is particularly important because the (mesoscopic) states $\as,\bs,\dots$ in the models considered here represent coarse-grained groups of microstates from an underlying physical system.   
For example, suppose that a given state comes in two variants (perhaps
$\bs,\bs'$) which differ in a way that is irrelevant for the
non-equilibrium driving force that controls the cycle affinity.
Fig.~\ref{fig:cycle_families} illustrates how this might appear in a
simple model: there are two cycles that proceed via $\bs,\bs'$ but
have the same affinity (because the driving force is blind to the
distinction between the states).  Since these two cycles have the same
affinity, they can be grouped into a family $\cal F$ and
(\ref{equ:BG-ave}-\ref{equ:P-KJ-bin}) still hold for the combined
counts. Moreover, the family could be extended by cycles that contain
arbitrary numbers of forward and backward jumps between $\bs$ and
$\bs'$, which would become relevant when the transition rates between
these to states are much faster than all other rates.

In this example, it is notable that the model may be coarse-grained
exactly by combining the states $\bs,\bs'$ into a single mesostate.
As such, the example illustrates that the results presented here are
consistent between different levels of coarse-graining. In fact, it is
generally sufficient to observe the system on a coarse-grained level,
as long the the coarse-graining does not mix cycles with different
affinities.  This mitigates the difficulty noted above, that the full
trajectory of a system must be observed in order to apply our results.

\subsection{Relation to fluctuation theorems}
\label{sec:fluct}

We have emphasised the connection between the results (\ref{equ:FT-simple}-\ref{equ:P-KJ-bin}) and fluctuation theorems~\cite{Gallavotti1995,Jarzynski1997,Crooks2000,AG,Seifert2012}.  As such, our derivations place the result (\ref{equ:BG}) of Biddle and Gunawardena~\cite{BG} in this context (under the restriction to non-revisiting cycles).
%The derivation of our main results puts the work of Biddle and Gunawardena~\cite{BG} in the context of fluctuation theorems~\cite{Gallavotti1995,Jarzynski1997,Crooks2000,AG,Seifert2012}. 
The central result that enables this analysis is~\eqref{eq:detailed_FT}, which can be regarded as an instance of the ``master fluctuation theorem'' of Ref.~\cite{Seifert2012}, employing our partial time-reversal~\eqref{eq:partial_timerev} as conjugate dynamics. 

Nonetheless, the results~(\ref{equ:FT-all},\ref{equ:P-KJ-bin}) differ from usual fluctuation theorems, as they involve the total count of cycle completions in either direction, as well as the net flux around a cycle, see also results for the traffic or frenesy~\cite{maes08,Maes2020}.

To connect to the more familiar case, note from \eqref{equ:P-KJ-bin} that $\tilde P_\tau(K,J)={\rm e}^{J{\cal A_C}}\tilde P_\tau(K,-J)$ and hence (summing both sides over $K$):
\beq
\frac{\tilde P_\tau(J)}{\tilde P_\tau(-J)}={\rm e}^{J{\cal A_C}} \; ,
\label{eq:FT_current}
\eeq 
similar to (\ref{equ:FT-all}).
This result is reminiscent of the fluctuation theorem for currents by Andrieux and Gaspard~\cite{AG}, but there are several important differences.

In particular, \eqref{eq:FT_current} concerns counting observables for cycle completions: {recall that $n_\CC$ and $\nR_\CC$ are the numbers of occurrences of specific sequences of states (for example $\CC=\as\bs\cs\as$ and $\CR=\as\cs\bs\as$) and $J_\CC$ is the difference between these numbers.  On the other hand, the result of~\cite{AG} concerns numbers of transitions between states (for example, one might consider a current defined as the difference between the number $\cs\to\as$ transitions and $\as\to\cs$ transitions).  From these numbers of transitions, one defines cycle currents} by an indirect method that involves a decomposition of steady-state current distributions in a basis that comes from Schnakenberg network theory~\cite{schn76}.

{We emphasise that the cycle currents in~\cite{AG} are distinct objects from the counting observables for cycle completions that we consider here.
{For example, consider the model of Fig.~\ref{fig:sketch1}: if we take  $\as\bs\cs\as$ and $\as\cs\ds\as$ as the fundamental cycles in the sense of~\cite{AG} (following the Schnakenberg formalism) then the trajectory $\as\bs\cs\ds\as$ would contribute $+1$ to each of the two \emph{cycle currents}~\cite{AG}.  However, the trajectory does not complete either of these cycles in the exact sequence given, so both there are no \emph{cycle completions} in the sense considered here (following~\cite{BG}).}

As a result of the indirect relationship between cycle currents and numbers of transitions, the fluctuation theorem of \cite{AG} appears as} a symmetry of the joint distribution of \emph{all} cycle currents.   Moreover, the Schnakenberg theory applies to steady-state currents, which means that the result of~\cite{AG} concerns the large-time limit of the current distribution.  The result of~\cite{AG} is a deep (and abstract) statement about the action of time-reversal on trajectories, and its implications for large deviations as $\tau\to\infty$.  On the other hand, it does not generally imply a fluctuation theorem for the (marginal) distributions of currents associated with individual cycles~\cite{mehl12,pole17,uhl18,kahl18}.

By contrast, (\ref{eq:FT_current}) is a much simpler result -- it applies for all $\tau$, for individual cycles.   The reason is that the cycle current is counted in a more direct way, by following the trajectory of the system throughout each instance of the cycle.  Since the initial and final states of the cycle are always equal, replacing an instance of $\CC$ by $\CR$ in trajectory $X$ has an effect on $P(X)$ that is simple, and does not affect other parts of the trajectory.

For the very special case of a unicyclic network -- and considering the family of cycles that include multiple forward and backward steps, as above -- the fluctuation theorem of \cite{AG} follows from (\ref{eq:detailed_FT}), in the long-time limit, 
{see also~\cite{Ge2008,Ge2012}}.  For multicyclic networks, the two results are distinct.   Given that fluctuations of cycle-counting observables contain new information, it may be that these results -- including that of Biddle and Gunawardena~\cite{BG} -- may prove useful for 
%alternative that overcomes the challenges inherent to 
thermodynamic 
inference, following~\cite{haya10,alem15}.
For that purpose, it is likely that inference based on families of cycles is more practical than counting instances of a specific cycle; for example, counting cycles within a family will typically result in larger observed numbers, improving the statistics.

\section{Large deviations as $\tau\to\infty$}
\label{sec:large}

Given the connection to fluctuation theorems~\cite{AG}, and that the original result of~\cite{BG} employed a large-time limit, it is useful to consider how cycle counting observables behave as $\tau\to\infty$.
%Given that $\nhat_\CC$ is the total number of completed cycles in $[0,\tau]$, 
One may expect by ergodicity that the cycle completion rate $\nhat_\CC/\tau$ converges to its steady state average as $\tau\to\infty$, which would be consistent with (\ref{equ:BG},\ref{equ:BG-ave}).  Large deviation theory provides a precise way to analyse this limit, and shows that this expectation is correct.  The relevant large-deviation methods can be found in~\cite{Lecomte2007,Touchette2009,Chetrite2015,Bertini2015b,Jack2020}, we outline the theory here. 

Define empirical time averages $\bar{k}=\hat{K}_\CC/\tau$ and $\bar\jmath = \hat{J}_\CC /\tau$: these are random (trajectory-dependent) quantities.  Their joint probability density behaves for long times as
\beq
P_\tau(\bar k,\bar\jmath) \approx \exp\left[ -\tobs\, {\cal I}(\bar k,\bar\jmath) \right],
\label{equ:ldp-kj}
\eeq
where ${\cal I}$ is the rate function, which is non-negative.  Such formulae are called large deviation principles -- they show that the typical values of $\bar k,\bar\jmath$ occur with probability one (hence ${\cal I}=0$), while other values have probabilities that become exponentially small as $\tobs\to\infty$.  
{They have been analysed for a different type of cycle counts in~\cite{Jia2016}.
% and there is also relevant work for large deviations of $m$th order Markov process~\cite{Chetrite-unpub}.}

The rate function may be characterised by the G\"artner-Ellis theorem as
\beq
{\cal I}(\bar k,\bar\jmath) = \sup_{s,\lambda} [ s\bar k + \lambda \bar\jmath  - \Psi(s,\lambda) ],
\eeq
where 
\beq
 \Psi(s,\lambda) = \lim_{\tobs\to\infty} \frac{1}{\tobs} \log \left\langle \exp(s\hat{K} + \lambda \hat{J}) \right\rangle 
 \label{equ:GE}
\eeq
is the scaled cumulant generating function (SCGF).  
Also, Varadhan's lemma states that
\beq
\Psi(s,\lambda)  = \sup_{\bar k,\bar\jmath} [ s\bar k + \lambda \bar\jmath  - {\cal I}(\bar k,\bar\jmath) ] \; .
\label{equ:varad}
\eeq
The marginal distribution for $\bar k$ obeys
\beq
P_\tau(\bar k) \approx \exp\left[ -\tobs\, {\cal I}_1(\bar k) \right]
\eeq
with ${\cal I}_1(\bar k) = \inf_{\bar\jmath} {\cal I}(\bar k,\bar\jmath)$, by the contraction principle.

A characterisation of $\Psi$ will be given below, as the largest eigenvalue of a matrix.  That analysis also ensures that the technical conditions required for \eqref{equ:ldp-kj} are satisfied, under our assumptions.
Before that, we explore how (\ref{equ:P-KJ-bin}) manifests itself in large deviations.

\subsection{Large deviations for non-revisiting cycles}

For non-revisiting cycles, we note that for large $K,J$ then \eqref{equ:P-KJ-bin} gives 
\beq
\frac{1}{\tobs} \log {P}_\tau(\bar k,\bar\jmath)  \approx \frac{1}{\tobs} \log {P}_\tau(\bar k)
- {\cal I}_2(\bar\jmath,\bar k,\Ac)
\eeq
where $\Ac$ is the cycle affinity and (by Stirling's approximation)
\begin{multline}
{\cal I}_2(\bar\jmath,\bar k,\Ac) =
 \bar k \log [ \cosh(\Ac/2) ]
-  \frac{\bar\jmath{\cal A}}{2 }
\\ +  \frac{\bar k + \bar\jmath}{2} \log \left(\bar k + \bar\jmath\right) 
+ \frac{\bar k - \bar\jmath}{2} \log \left(\bar k - \bar\jmath\right) \; .
\end{multline}
Then \eqref{equ:ldp-kj} implies that the rate function is
\beq
{\cal I}(\bar k,\bar\jmath) = {\cal I}_1(\bar k)  +{\cal I}_2(\bar\jmath,\bar k,\Ac) \; .
\eeq
The function ${\cal I}_2$ 
is closely related to the rate function for the time-averaged current $\bar\jmath$ of a biased random walk, which is related in turn to the binomial structure of \eqref{equ:P-KJ-bin}.

Now define $g(\lambda,\Ac) = {\bar k}^{-1} \sup_{\bar\jmath} [ \lambda\bar\jmath - {\cal I}_2(\bar\jmath,\bar k,\Ac)]$ and observe that
\beq
g(\lambda,\Ac) = \log \frac{\cosh(\lambda+\Ac/2)}{\cosh(\Ac/2)} \; .
\label{eq:gfunc}
\eeq
(It is important that this object does not depend on $\bar k$: while this is not obvious from its definition, it follows from the relationship of ${\cal I}_2$ to a random walk.)
Also define $\Psi_1(s) = \Psi(s,0)$ as the SCGF for $\bar k$.
Then by (\ref{equ:varad}) the (joint) SCGF $\Psi$ has the simple form
\beq
\Psi(s,\lambda) = \Psi_1(s+g(\lambda,\Ac)) \; .
\label{equ:Psi-Psi1}
\eeq
[The right hand side is the function $\Psi_1$ evaluated at the point $s+g(\lambda,\Ac)$.] The function $g$ is symmetric with respect to $\lambda=-\Ac/2$. This symmetry gets inherited by the SCGF $\Psi(s,\lambda)$, where it reflects the fluctuation relation~\eqref{eq:FT_current}.

The expression \eqref{equ:Psi-Psi1} is simple in that the function $g$ depends on system parameters only through the affinity $\Ac$, while the effects of all other properties of the system are encoded in a single function $\Psi_1$.  Similarly in (\ref{equ:P-KJ-bin}), the conditional distribution of $J$ (given $K$) is binomial and depends only on $\Ac$, but the distribution {$\tilde{P}_{\tau}(K)$} depends in a non-trivial way on all system parameters.  In this sense, (\ref{equ:Psi-Psi1}) is the consequence for large deviations of the detailed result (\ref{equ:P-KJ-bin}) for finite times. 

 Fig.~\ref{fig:Psi} illustrates (\ref{equ:Psi-Psi1}) in the simple example of Fig.~\ref{fig:sketch1}, for the cycle $\CC=\as\bs\cs\as$.  The numerical computation of the SCGFs was performed using the method described in Appendix~\ref{app:computation}.  
 Contours of the SCGF are the lines $s=g(\lambda,\Ac)$: we show results for two sets of system parameters, which lead to the same value of $\Ac$; hence the contour lines are the same in both cases, although the corresponding values of $\Psi$ differ {by an order of magnitude}.  {Hence, the fact that these figures appear similar (despite the different model parameters) shows that the theoretical result (\ref{equ:Psi-Psi1}) does indeed apply.  This is a direct consequence (at the level of large deviations) of the binomial distribution of (\ref{equ:P-KJ-bin}), which is the key result from which the other fluctuation properties are derived, in this work.}

\begin{figure}
  \centering
  \includegraphics[width=\linewidth]{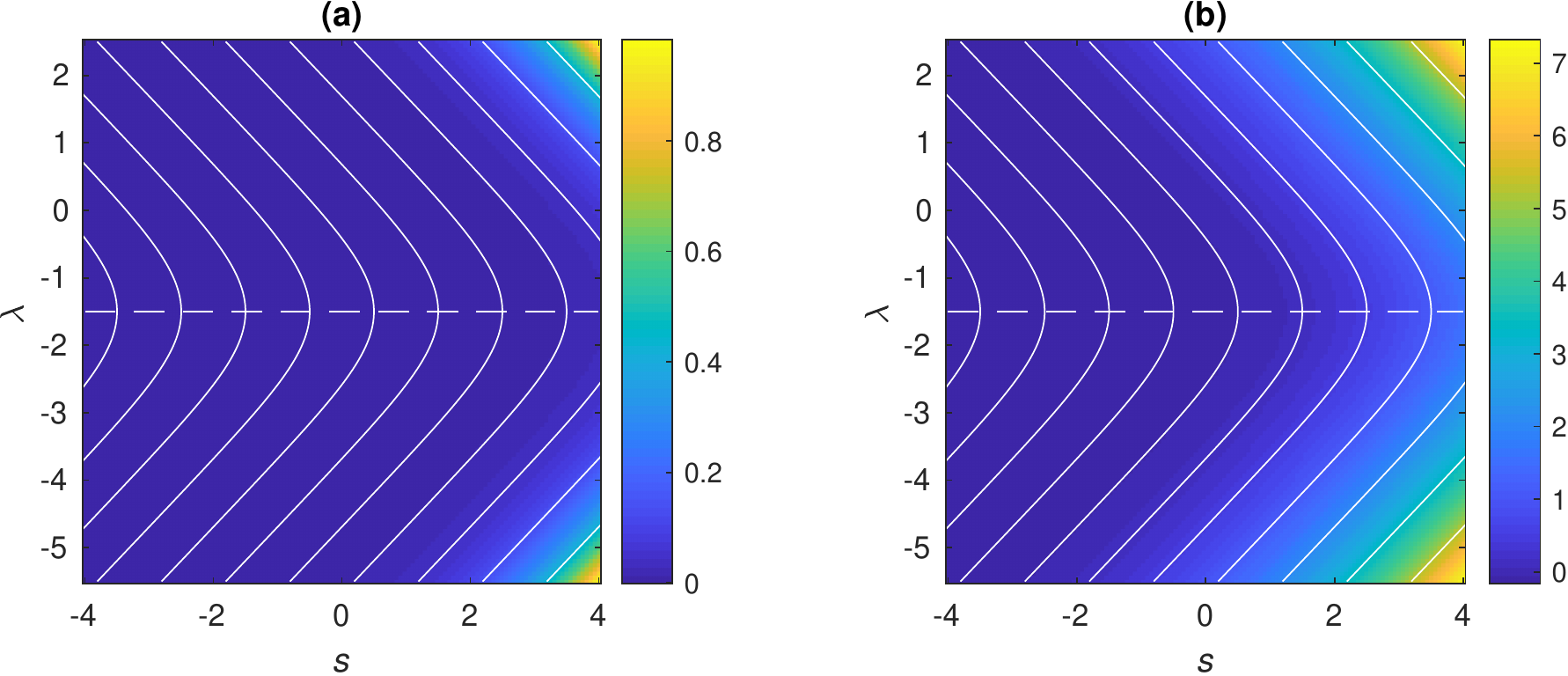}
  \caption{SCGF $\Psi(s,\lambda)$ (colour coded) for the current and traffic of the cycle $\CC=\as\bs\cs\as$ in Fig.~\ref{fig:sketch1}. (a) and (b) differ in the choice of rates, but the affinity ${\cal A_C}=3$ is fixed. Lines of constant value of $s+g$ are shown in solid white. They prescribe the overall shape of the SCGF, including the symmetry with respect to $\lambda=-{\cal A_C}/2$ (dashed white line) corresponding to the fluctuation symmetry~\eqref{eq:FT_current}.  Parameters: (a) $w(\as\to \bs)=w(\bs\to \cs)=w(\cs\to \as)=0.5$, $w(\cs\to \as)=w(\ds\to \as)=2$, $w(\bs\to \as)=w(\as\to \cs)=0.5{\rm e}^{-2}$, $w(\cs\to \bs)=0.5{\rm e}$, $w(\ds\to \cs)=w(\as\to \ds)=2{\rm e}^{2}$; (b) $w(\as\to \bs)=w(\bs\to \cs)=w(\cs\to \as)=1$, $w(\cs\to \ds)=w(\ds\to \as)=w(\as\to \ds)=0.1$, $w(\bs\to \as)=w(\cs\to \bs)=w(\as\to \cs)={\rm e}^{-1}$, $w(\ds\to \cs)=0.1{\rm e}^{2}$.  }
  \label{fig:Psi}
\end{figure}

\subsection{Large deviations for words and cycles}% via $m$th order Markov processes}
\label{sec:mth_order_ldf}

This section outlines a general method for analysis of large deviations of cycle counts.  
This establishes that \eqref{equ:ldp-kj} does indeed hold, and provides a method for computation of SCGFs. 
Similar methods are used for analysis of word-counting in DNA sequence analysis~\cite{Lothaire2005} and in the statistics of repeated measurements~\cite{Horssen2015}, see also~\cite{Jia2016}.

{Similar SCGFs to (\ref{equ:GE}) appear when considering large deviations of the number of transitions between discrete states of Markov models -- for example one might define $\hat{n}_2$ as the number of transitions $\as\to\bs$ and $\hat{n}_2^{\rm R}$ as the number of transitions $\bs\to\as$.  Then consider (\ref{equ:GE}) with $\hat{K},\hat{J}$ replaced by $\hat{K}_2=\hat{n}_2+\hat{n}_2^{\rm R}$ and $\hat{J}_2=\hat{n}_2-\hat{n}_2^{\rm R}$ respectively.  The resulting SCGF can be obtained by established methods \cite{Lecomte2007,Touchette2009,Chetrite2015,Bertini2015b,Jack2020} as the largest eigenvalue of a particular matrix that is called the \emph{tilted generator}.}

{However, the established methodology is not applicable in the current setting because $\hat{n}_{\CC}$ is not obtained by counting transitions between pairs of states (nor by considering state occupancies) -- it requires that we count occurrences of specific \emph{words}.
The solution} is to expand the state space of the original system to obtain an \emph{extended system} in which each state is a word of length $m$.   We illustrate this with the case $m=3$.  Suppose that the (original) system is in state $\cs$ and the previous two states visited were $\as,\bs$, in that order.  Then the state of the extended system is the 3-letter word $\as\bs\cs$.  If the original process now makes a transition to $\as$ then the extended system makes a transition to $\bs\cs\as$.   (After the transition, the state is $\as$ and the previous two states were $\bs,\cs$.)  This example is useful because this transition $\as\bs\cs\to \bs\cs\as$ in the extended system corresponds exactly with a completed cycle in the original system.  
{In other words, the problem of word-counting in the original model is reduced to a problem of counting transitions between states of the extended model.  Since the extended model is still Markovian, established methods can then be used to compute the statistics of the relevant transitions, see below.}
%This extended system is an example of an $m$th order Markov process~\cite{Lothaire2005}.

As a technical remark: this construction provides a mapping between trajectories of the original and extended systems, so that cycle counts of the original system can be inferred from the extended one.  However, the initial $m-1$ states of a trajectory of the extended system are not fully-determined by a trajectory of the original system.  This issue can cause some ambiguity in cycle counts; but the problem can easily be rectified to obtain a one-to-one mapping of trajectories. Since the behavior of the first few states will not affect large-deviation analysis, we do not discuss this aspect.

To define more precisely the extended system,  focus on a specific cycle and take $m=m_{\CC}$.  Each state of the extended system is an $m$-letter word (for example $\as\bs\cs$ or $\bs\cs\as$), we denote these words by $u,v,\dots$.  The transition rates of the extended system are denoted by $W(u\to v)$.   The rate $W(u\to v)$ is non-zero only if the first $m-1$ letters of word $u$ are the same as the first $m-1$ letters of $v$.  In this case $W(u\to v)=w(u_{\rm f}\to v_{\rm f})$ where $u_{\rm f}$ is the final letter of word $u$, and similarly $v_{\rm f}$ (recall that $w$ indicates is a transition rate of the original system).
 One sees that construction of this extended system is a straightforward exercise, although it can be tedious because the number of states grows quickly with the word length and the number of states in the original system. For practical purposes, a milder extension of the state space is sufficient to establish specific results for cycle counts, see Appendix~\ref{app:computation}.

Now write $u_\CC$ for the first $m_\CC$ letters of $\CC$ and $v_\CC$ for its last $m_\CC$ letters.  (In the example $\CC=\as\bs\cs\as$ then $u_\CC=\as\bs\cs$ and $v_\CC=\bs\cs\as$.)
Then completion of cycle $\CC$ corresponds to a transition $u_\CC\to v_\CC$ in the extended system, that is
\beq
\nhat_\CC(X) = N_{u_\CC\to v_\CC}(X)
\label{equ:n-N}
\eeq
where $N_{u\to v}(X)$ is the number of transitions $u\to v$ in trajectory $X$ of the extended system. 
In the same way,
\beq
\nRhat_\CC(X) = N_{u_\CR\to v_\CR}(X)
\eeq
where $u_\CR$ indicates $u_{\CR}$, the first $m_\CC$ letters of $\CR$, and similarly $v_\CR$.

The extended system is itself Markovian, so standard methods can be used to analyse its large deviations.   In particular, a method for counting transitions $N(u\to v)$ between states is well-established~\cite{Lecomte2007,Touchette2009,Chetrite2015,Bertini2015b,Jack2020}, {we give an outline, with details in Appendix~\ref{app:tilt}.}
The master equation of the extended system takes the standard form
\beq
\frac{\partial}{\partial t}P(u,t) = \sum_{v(\neq u)} \left[ P(v,t) W(v\to u) - P(u,t) W(u\to v) \right] .
\label{equ:master-uv}
\eeq
Now define a matrix ${\mathbb W}$ with off-diagonal elements $[{\mathbb W}^0]_{vu}=W(u\to v)$ and diagonal elements $[{\mathbb W}^0]_{uu}=-\sum_v W(u\to v)$.  Then the master equation is $\partial_t P = {\mathbb W}^0P$, where $P$ is interpreted as a vector with elements $P(u)$.  The SCGF can be obtained as the largest eigenvalue of the (``tilted'') matrix
\beq
{\mathbb W}(s,\lambda)  = {\mathbb W}^0 + {\mathbb V}(s,\lambda),
\label{equ:tilted_matrix}
\eeq
where ${\mathbb V}(s,\lambda)$ has only two elements that are non-zero:
\begin{align}
[{\mathbb V}(s,\lambda)]_{v_\CC,u_\CC} & = ({\rm e}^{s+\lambda}-1) [{\mathbb W}^0]_{v_\CC,u_\CC} \; ,
\nonumber\\
[{\mathbb V}(s,\lambda)]_{v_\CR,u_\CR} & = ({\rm e}^{s-\lambda}-1) [{\mathbb W}^0]_{v_\CR,u_\CR} \; .
\end{align}

To establish that (\ref{equ:ldp-kj},\ref{equ:GE}) hold, a few technical conditions are required on $\mathbb{W}(s,\lambda)$.  Note that the extended process is Markov with a finite state space.  In this case it is sufficient for it to have a unique steady state, which must hold if the original system is irreducible, as assumed above.  Hence one has a large deviation result of the form (\ref{equ:ldp-kj}).

Note that this construction is fully general, there was no assumption of non-revisiting cycles. 
If one does assume that $\CC$ is non-revisiting, the largest eigenvalue of ${\mathbb W}(s,\lambda) $ must be of the form \eqref{equ:Psi-Psi1}.  An explicit derivation of this result is deferred to future work, which might also consider how large-deviation properties can be computed from the representation of the cycle-counting problem as a kind of renewal process via (\ref{equ:renew-path}), and what generalizations of the fluctuation theorems are possible for revisiting cycles.

\section{Conclusion}
\label{sec:conc}

We have analysed the joint distribution of cycle counts for forward and backward instances of a cycle $\CC$ in a discrete Markov process, as commonly used for analysis of non-equilibrium systems.  The distribution is naturally characterised in terms of the cycle current $J$ and the total count $K$.  For non-revisiting cycles (which are those of primary physical relevance), the central result is (\ref{equ:P-KJ-bin}), which shows that the conditional distribution of $J$ given $K$ is binomial and the only relevant parameter is the affinity.  This shows that the conditional distribution of $J$ is universal, with the affinity as its only parameter, while the distribution of $K$ is free and depends on all system details.

For practical purposes, we point to (\ref{equ:BG-ave}) as a finite-time generalisation of (\ref{equ:BG}), which might be useful as a way to infer affinities, as proposed in~\cite{BG}. {The counting of instances of cycle families rather than individual cycles, as discussed in Sec.~\ref{sec:generalisation}, might also help to improve this method.}

We have also explained how large deviation theory can be applied to cycle counts.  In particular, they do obey a large-deviation principle, whose properties can be computed from the extended system described here, by solving an eigenvalue problem.

These results suggest that further interesting structure may be present in distributions of cycle counts, either by analysis of the extended system, or by considering joint distributions of counts across more than one cycle.  We look forward to future work in this direction.

\begin{acknowledgments}
We thank Jeremy Gunawardena and John Biddle for helpful discussions. 
J.G. acknowledges funding from the Royal Society under grant No. RP17002.
This work was funded in part by the European Research Council under the EU's Horizon 2020 Program, Grant No. 740269.  
\end{acknowledgments}

\begin{appendix}

\section{Derivation of (\ref{equ:PC},\ref{equ:F-fin-R})}

{
For completeness, we derive (\ref{equ:PC},\ref{equ:F-fin-R}), starting from (\ref{eq:pathweight}).  First note that for {any trajectory $X$} starting at time $t$ and ending at time $\tau$, the analogue of  the path weight (\ref{eq:pathweight}) can be written as 
\begin{align}
P[X]=p(x_0,t)&\left[\prod_{i=0}^{M-1} w(x_i\to x_{i+1}){\rm e}^{-r(x_i) \Delta_i }\right]\nonumber\\&\qquad\times{\rm e}^{-r(X_M) (\tau-t_{M})}
\label{eq:pathweight-t}
\end{align}
where $\Delta_i=t_{i+1}-t_i$ is the sojourn time in state $x_i$ and $p(x_0,t)$ is the probability distribution of the initial state (at time $t$).  The states $(x_0,x_1,\dots,x_M)$ and times $(t_1,\dots,t_M)$ are indexed from time $t$, note also $t_0=t$.
This distribution $P[X]$ is normalised in the sense that
\beq
1 = \sum_{M=0}^\infty \sum_{x_0\dots x_M} \int_0^\tau dt_1 \int_{t_1}^\tau dt_2 \dots \int_{t_{M-1}}^\tau dt_M P[X] \; .
\eeq

Using this distribution, and given a cycle $\CC$, we compute the probability $P_\CC(t,\tau)\delta t$ of the following event: the trajectory has $t_{m_\CC}<\tau$ (from which it follows that $M\geq m_\CC$); also $(x_0,x_1,\dots,x_{m_\CC})=(\CC_1,\CC_2,\dots,\CC_{m_\CC},\CC_1)$, and  $t_1\in[t,t+\delta t]$. 
We use (\ref{eq:pathweight-t}) and sum over those $x_k$ with $k>m_\CC$, and integrate all the $\Delta_i$, to obtain [at leading order in $\delta t$]:
\begin{multline}
P_\CC(t,\tau) \delta t = p(\CC_1,t) w(\CC_1\to\CC_2)  \delta t  \prod_{j=2}^{m_\CC} w(\CC_j\to \CC_{j+1})
 \\  \times
\int d\Delta_2 \dots d\Delta_{m_\CC}  \Theta\left( \tau - t - \sum_{j=2}^{m_\CC} \Delta_j\right) {\rm e}^{-\sum_{j=2}^{m_\CC} r(\CC_j) \Delta_j }.
\label{equ:PC-intermediate}
\end{multline}
Here $\Theta$ is the Heaviside (step) function; the $\Delta_j$ are integrated over $[0,\infty)$; we used that the integral for $\Delta_0$ runs over $[t,t+\delta t]$, which yields the factor $\delta t$.  Eq.~(\ref{equ:PC-intermediate}) coincides with (\ref{equ:PC}) if we identify
\begin{multline}
F_{\rm fin}(\tau-t,\CC) = \int d\Delta_2 \dots d\Delta_{m_\CC} \Theta\left( \tau - t - \sum_{j=2}^{m_\CC}\Delta_j\right)
\\
\times \prod_{j=2}^{m_\CC} \left[ r(\CC_j) {\rm e}^{- r(\CC_j) \Delta_j }  \right] \; .
\label{equ:F-fin-explicit}
\end{multline}
To interpret this result, we identify $\Delta_{\rm tot} = \sum_{j=2}^{m_\CC} \Delta_j$ as the sum of $m_\CC-1$ exponential random variables with means $r(\CC_2)^{-1},\dots,r(\CC_{m_\CC})^{-1}$.
As advertised in the main text, the result (\ref{equ:F-fin-explicit}) is simply the probability that this $\Delta_{\rm tot}$ is less than $\tau-t$.  For any given $m_\CC$, the integrals can be performed, but we retain here the integral form, which shows the structure of the result.  In particular, it is clear from (\ref{equ:F-fin-explicit}) that (\ref{equ:F-fin-R}) holds,
%\beq
%F_{\rm fin}(\tau-t,\CC)  = F_{\rm fin}(\tau-t,\CR) 
%\eeq
because $\CR$ contains the same states as $\CC$ (only the order is reversed), and {the factor $F_{\rm fin}(\tau-t,\CC)$} from (\ref{equ:F-fin-explicit})  is invariant under permutation of the states $\CC_2,\dots,\CC_{m_{\CC}}$ within the cycle $\CC$.
}

\section{Connection to renewal theory}
\label{app:renew}

\subsection{Alternative derivation of (\ref{equ:P-KJ-bin}) by renewal theory}
\label{app:alt_proof}

The results (\ref{equ:FT-simple}-\ref{equ:P-KJ-bin}) for non-revisiting cycles can also be proven using a methodology similar to renewal processes.  We include this analysis for completeness, and because the results provide additional information on the statistics of cycle completions, that may be useful for future work.

  Suppose that  the cycle $\CC$ of interest starts in state $\as$.  Any trajectory $X$ can be decomposed into several pieces as in 
Fig.\ \ref{fig:sketch3}: an initial transient before the first visit to $\as$, the time periods spent in $\as$, the complete cycles between visits to $\as$, and a final period between the last visit to $\as$ and the end of the trajectory at time $\tau$.   Moreover, for any cycle $\CC$, one can classify the complete cycles as instances of either $\CC$, or $\CR$, or some other cycle.  

\newcommand{\OO}{\mathcal{O}}

Hence, any trajectory $X$ can be associated to a reduced trajectory $Y$, which is
is characterised by the sequences of arrival and departure times to/from $\as$ and the sequence of cycle types, for example 
\beq
{\cal S}_Y = ( \CC,\OO,\CC,\CR,\OO,\dots)
\eeq
where $\OO$ denotes any cycle other than $\CC,\CR$.  (Separate occurrences of ${\cal O}$ may indicate different cycles.)  It is assumed that the cycle begins and ends with generic words that are  indicated by $\cal W$ in Fig.~\ref{fig:sketch3}, these are not included in ${\cal S}_Y$.  If the trajectory starts or ends in $\as$ then one or both of the ${\cal W}$s will have zero length.  Compared with (\ref{equ:seq-example}), this ${\cal S}_Y$ is different in that it includes a separate element for every departure from $\as$, not only those departures that lead to cycles $\CC$ or $\CR$.  Similarly, we use $t^{\rm dep}_i$ and $t^{\rm arr}_i$ in this Section to indicate the times of (all) departures/arrivals from/to $\as$.

The mapping from $X$ to the reduced trajectory $Y$ is many-to-one because the times for transitions inside the cycles are not preserved, and nor are the sequences of states inside  the generic cycles/words ${\cal O},{\cal W}$.  In the following, we consider the probabilities of the reduced trajectories $Y$, which are obtained by integrating over all possible trajectories $X$ that reduce to $Y$.  One sees that $\nhat_\CC(X)$ is the number of occurrences of $\CC$ in ${\cal S}_Y$ (and similarly for $\CR$), so the full statistics of $\nhat_\CC,\nRhat_\CC$ can be computed from the statistics of the reduced trajectories $Y$.

The probabilities of the reduced trajectories have several useful properties, which are summarized here, with extra detail in Appendix~\ref{app:red}.  First, the times between each arrival in $\as$ and the next departure are all independent, they are exponentially distributed with mean $r(A)^{-1}$.  Second, on departure from $\as$ at time $t_{\rm dep}$, the subsequent behaviour is Markovian (independent of the previous history), as also occurs in renewal processes.   The probability that any departure from $\as$ leads to a complete cycle $\CC$ is [similar to \eqref{equ:PC}]
\beq
P^{\rm ren}_\CC(\tau-t_{\rm dep}) =  \frac{w(\as \to \CC_2)}{ r(\as) } p_{\rm seq}(\CC) F_{\rm fin}(\tau-t_{\rm dep},\CC) \; .
\eeq 
%The factor $F_{\rm fin}$ accounts for the probability that the cycle is not completed during the trajectory (that is, before time $\tau$).  
Also, given that such a cycle is completed, the time $t_{\rm arr}$ for the next arrival in $A$ (which is the end time of the cycle) has cumulative distribution function 
\beq
{\cal P}(t_{\rm arr}  < t_{\rm dep}+\Delta t \,|\, \CC) = \frac{ F_{\rm fin}(\Delta t,\CC) }{ F_{\rm fin}(\tau-t_{\rm dep},\CC) }   \; .
\eeq

\begin{figure}
\includegraphics[width=8.5cm]{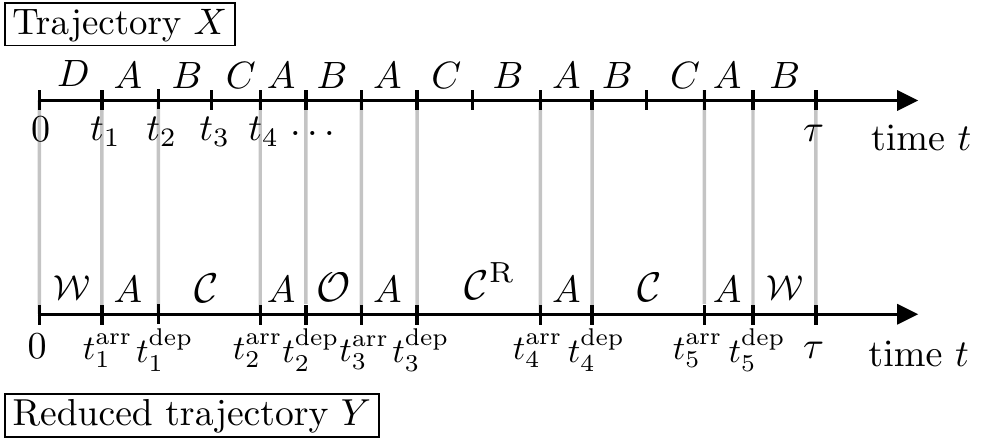}
\caption{A trajectory $X$ of the four-state system in Fig.~\ref{fig:sketch1}, and the corresponding reduced trajectory $Y$.
The reduced trajectory keeps track of the arrival and departure times for state $A$, and on whether excursions from $A$ are instances of $\CC$ or $\CR$ or some other cycle (indicated as $\cal O$).  The initial and final parts of the trajectory consist of generic words (indicated by $\cal W$).  This reduced trajectory has ${\cal S}_Y = (\CC,{\cal O},\CR,\CC)$}
\label{fig:sketch3}
\end{figure}

Hence, given that the system departs from $\as$ at time $t_{\rm dep}$, the probability density that it completes an instance of cycle $\CC$ and returns to $\as$ a time $\Delta t$ later is
\beq % \begin{multline}
f_\CC(\Delta t)  = \frac{w(A \to \CC_2)}{ r(A) } p_{\rm seq}(\CC) 
%\\ \times 
\frac{\partial}{\partial (\Delta t)} F_{\rm fin}(\Delta t,\CC)   \; . 
\label{equ:fC}
\eeq %\end{multline}
Using \eqref{equ:F-fin-R} and \eqref{equ:aff},
the corresponding quantity for $\CR$ is
\beq
f_{\CR}(\Delta t)   = {\rm e}^{-\Ac_\CC} f_\CC(\Delta t)  \; .
\label{equ:f-fR}
\eeq
Following similar arguments, a formula is available for the probability of any reduced trajectory.  This is given in Appendix~\ref{app:red}.

Notwithstanding that derivation, an important fact is already apparent from \eqref{equ:f-fR}:  Given any reduced trajectory $Y$, one may obtain a new trajectory $Y'$ by replacing any instance of $\CC$ in ${\cal S}_Y$ by $\CR$ (keeping all other aspects of the trajectory fixed).  The resulting trajectory probabilities are related as
\beq
{\cal P}(Y') =  {\cal P}(Y) {\rm e}^{-\Ac_\CC} 
\label{equ:PY-PY'}
\eeq
[See also Appendix~\ref{app:red}, and note that this is analogous to (\ref{eq:detailed_FT}).]

Recalling from \eqref{equ:JK} that $\hat{K}$ is the total number of instances of $\CC$ and $\CR$, one may define a set $\Lambda$ containing $2^{\hat{K}}$ (reduced) trajectories, formed by all possible replacements of $\CC$ by $\CR$, and vice versa.  Then the conditional probability of trajectory $Y$  within this set is 
\beq
{\cal P}(Y|\Lambda) = \frac{ {\rm e}^{-\Ac_\CC \nRhat_\CC(Y) } }{ ( 1 + {\rm e}^{-\Ac_\CC} )^{K_\Lambda} } 
\label{equ:PY-lambda}
\eeq
where $K_\Lambda$ is the value of $\hat{K}$ for all trajectories in $\Lambda$.
These trajectories have different values of $\hat{J}$; the number of trajectories with any given value is a binomial coefficient.  Hence [using \eqref{equ:JK}] the conditional distribution of $\hat J$ is
\beq
P(  J|\Lambda) = 
 \begin{pmatrix}   K_\Lambda \\ \frac12(K_\Lambda+  J) \end{pmatrix} 
\frac{\exp({ J\Ac/2}) } 
{  [2\cosh(\Ac/2)]^{K_\Lambda} } \; .
\label{equ:P-J-Lambda}
\eeq
Finally, the distribution $\tilde{P}_\tau(K,J)$ of (\ref{equ:P-KJ-bin}) can be obtained by conditional probability as $\tilde{P}_\tau(K,J) = \sum_{\Lambda|K} P(J|\Lambda) P(\Lambda)$ where the sum (which might alternatively be expressed as an integral) is over all sets $\Lambda$ with $K_\Lambda=K$, and $P(\Lambda)$ is the probability that a random trajectory is in the set $\Lambda$.   This yields (\ref{equ:P-KJ-bin}).

\subsection{Probabilities of reduced trajectories}
\label{app:red}

We derive the probability of a reduced trajectory $Y$, whose definition is illustrated in Fig.~\ref{fig:sketch3}.  On each visit to state $\as$, the system loses all memory of its previous history: this is a renewal.  It follows that the probability of trajectory $Y$ is given by a product of terms, one from each of its components.  Denote the number of visits to $\as$ by ${\cal N}$, this is a random quantity but we do not write any hats, to lighten the notation.  Hence $Y$ is specified by ${\cal N}$ arrival times and ${\cal N}$ departure times, and the ${\cal N}-1$ elements of ${\cal S}_Y$.  
If the trajectory starts in $\as$ then we take $t^{\rm arr}_1=0$ and if it ends in $\as$ then $t^{\rm dep}_{\cal N}=\tau$. 
The initial condition of the system is given by a distribution $p_{\rm ini}$ over its states (it is not assumed that $p_{\rm ini}$ corresponds to the steady state).  

The first contribution to the trajectory probability comes from the transient period before the first visit to $\as$, it is a probability density for $t^{\rm arr}_1$, which we write as $f_{\rm beg}(t^{\rm arr}_1)$.  This probability has two contributions, the first is $p_{\rm ini}(\as) \delta(t^{\rm arr}_1)$ because the system may start in $\as$.  The second is the probability density that the system first reaches $\as$ at time $t^{\rm arr}_1$.  This distribution can be computed if necessary, for the purposes of this work it is sufficient that $f_{\rm beg}(t^{\rm arr}_1)$ exists, but the specific form is not required.

The next contribution comes from the visits to $\as$.  After each arrival, the system stays in $\as$ for a time $t_{\rm dep}-t_{\rm arr}$ whose probability density is 
\beq
f_\as(t_{\rm dep}-t_{\rm arr}) = r(\as) {\rm e}^{-(t_{\rm dep}-t_{\rm arr}) r(\as)}.
\eeq
The next contribution comes from completed cycles.  On leaving $\as$ at time $t_{\rm dep}$, the probability to complete a cycle $\CC$ and return a time $\Delta t$ later is $f_\CC(\Delta t|t_{\rm dep}) $ as given in \eqref{equ:fC}.  A similar expression holds for cycle $\CR$, see \eqref{equ:f-fR}.  One must also consider the probability density to return to $\as$ by a different cycle (neither $\CC$ or $\CR)$ after time $\Delta t$, which is denoted by $f_{\cal O}(\Delta t|t_{\rm dep})$.   The precise form of this function is not needed for the current purpose, only that it is well-defined (similar to $f_{\rm beg}$).
Still, if one considers very long trajectories, a system that departs from $\as$ must eventually return to it, from which one deduces the normalization constraint
\beq
\int_0^\infty {\rm d}t \left[ f_{\CC}(t)  + f_{\CR}(t)  + f_{\cal O}(t) \right] = 1  \; .
\eeq

Finally one must consider the contribution to the trajectory probability from the final component, between the last departure from $\as$ and time $\tau$.   This is denoted by $p_{\rm end}(\tau-t^{\rm dep}_{\cal N})$.  The form of this contribution depends on whether the system ends the trajectory in state $\as$ (so $\tau=t^{\rm dep}_{\cal N}$) or not.  In the latter case, $p_{\rm end}(t)$ is the probability that a system departing from state $\as$ does not return to it within time $t$.  In the case $\tau=t^{\rm dep}_{\cal N}$ then $p_{\rm end}$ has a contribution $r(\as)^{-1}\delta(t)$, this factor combines with the $f_\as$ contributions to the trajectory probability to ensure that the distribution of times spent in $\as$ is correctly accounted for.

Combining all these ingredients, the probability density for the reduced trajectory $Y$ is
\begin{multline}
{\cal P}(Y) = f_{\rm beg}(t^{\rm arr}_1) \left[ \prod_{i=1}^{\cal N} f_\as(t^{\rm dep}_{i} - t^{\rm arr}_{i}) \right]
f_{\rm end}(\tau-t^{\rm dep}_{{\cal N}})
\\ \times \prod_{i=1}^{{\cal N}-1} f_{({\cal S}_Y)_i}(t^{\rm arr}_{i+1} - t^{\rm dep}_{i})
\label{equ:renew-path}
\end{multline}
Here $f_{({\cal S}_Y)_i}$ is one of $f_{\CC},f_{\CR},f_{\cal O}$, according to which kind of cycle appears in the $i$th element of ${\cal S}_Y$.

From this final result one may directly check \eqref{equ:PY-PY'}, because the only change on replacing an instance of $\CC$ in $Y$ by $\CR$ is to exchange a factor of $f_{\CC}(t^{\rm arr}_{i+1} - t^{\rm dep}_{i})$ for $f_{\CR}(t^{\rm arr}_{i+1} - t^{\rm dep}_{i})$.  Using this with \eqref{equ:f-fR} yields \eqref{equ:PY-PY'}.

\section{Large deviation computation}
\subsection{SCGF for large deviations for cycle counts}
\label{app:tilt}

{We outline the derivation of the SCGF from (\ref{equ:GE}) as the largest eigenvalue of the matrix $\mathbb{W}(s,\lambda)$ in (\ref{equ:tilted_matrix}).
We also explain why this SCGF cannot be derived by applying a ``standard'' tilting method to the original system.

Following (for example)~\cite{Lecomte2007,Garrahan2009}, we generalise the probability $P(u,t)$ from (\ref{equ:master-uv}) by defining $P(u,n_\CC,n_\CR,t)$ as the probability for the extended system to be in state $u$ at time $t$, having made $n_\CC$ completions of cycle $\CC$ and $n_\CR$ completions of $\CR$.  It is crucial that this $P$ obeys its own master equation:
\begin{multline}
\frac{\partial}{\partial t}P(u,n_\CC,n_\CR,t) = \sum_{v(\neq u)}  P(v,n_\CC,n_\CR,t) W(v\to u) 
\\ - \sum_{v(\neq u)}   P(u,n_\CC,n_\CR,t) W(u\to v) 
\\ + [ P(u_\CC,n_\CC-1,n_\CR,t) -P(u_\CC,n_\CC,n_\CR,t)  ]W(u_\CC\to v_\CC) \delta_{u,v_\CC}
\\ + [ P(u_\CC,n_\CC,n_\CR-1,t) -P(u_\CC,n_\CC,n_\CR,t)  ]W(u_\CR\to v_\CR) \delta_{u,v_\CR}
 .
\label{equ:master-uvn}
\end{multline}
where the 3rd and 4th lines account for the fact that transitions $u_\CC\to v_\CC$ and $u_\CR\to v_\CR$ correspond to cycle completion events, in which the value of either $n_\CC$ or $n_\CR$ changes.  [Recall Eq.~(\ref{equ:n-N}).]
Now define 
\beq
\tilde{P}(u,s,\lambda,t) = \sum_{n_\CC,n_\CR} P(u,n_\CC,n_\CR,t) {\rm e}^{ (s+\lambda)n_\CC + (s-\lambda)n_\CR }
\eeq
(the sums run from $0$ to $\infty$).  Note that $P$ is a normalised probability distribution over $(u,n_\CC,n_\CR)$ but $\tilde{P}$ is not normalised. 
Then by (\ref{equ:master-uvn}) one has
\beq
\frac{\partial}{\partial t} \tilde{P}(u,s,\lambda,t)  = \sum_{v} {[\mathbb{W}(s,\lambda)]_{u,v}}  \tilde{P}(v,s,\lambda,t) 
\label{equ:tilted-wuv}
\eeq
where the matrix ${\mathbb{W}(s,\lambda)}$ is defined in (\ref{equ:tilted_matrix}).  This equation corresponds to $\partial_t \tilde P = {\mathbb{W}_{u,v}(s,\lambda)} \tilde P$ from which one sees that the long-time behaviour of $\tilde{P}$ is dominated by the largest eigenvalue of $ {\mathbb{W}(s,\lambda)}$, that is $\tilde{P}(u,s,\lambda,t) \simeq \tilde{P}_\infty(u,s,\lambda) {\rm e}^{t\Psi(s,\lambda)}$ for large times, where $\Psi(s,\lambda)$ is the largest eigenvalue.  Summing over $u$, one may establish that the SCGF (\ref{equ:GE}) coincides with this $\Psi$, as in~\cite{Lecomte2007,Garrahan2009}.

To see that this method requires the extended system, note that (\ref{equ:master-uvn}) describes a Markovian dynamics for the evolution of $(u,n_\CC,n_\CR)$, where $u$ is the state of the extended system.  By contrast, if one considers the original system (whose state is $x$), the evolution of $(x,n_\CC,n_\CR)$ is not Markovian: the probability of an event where $n_\CC$ increases depends on the history of recently-visited states, and not only on the current state $x$.  (Specifically, $n_\CC$ can only increase if the current state is $\CC_{m_\CC}$ and the previous $m_\CC-1$ states were $\CC_1,\dots,\CC_{m_\CC-1}$.)   As a result, the recipe given here -- which connects SCGFs to eigenvalues -- is only applicable at the level of the extended system.

In fact, the evolution of the state $(x,n_\CC,n_\CR)$ is an example of an $m$th order Markov process (we refer to \cite{Lothaire2005} for applications of such models to word counting in discrete time, and to \cite{Chetrite-unpub} for a discussion of large deviations in $m$th order Markov processes).
}

\subsection{Practical calculation of large deviations for cycle counts}
\label{app:computation}

As discussed in Sec.~\ref{sec:mth_order_ldf}, the SCGF $\Psi$ for cycle counts can be characterised as the the largest eigenvalue of a matrix, which is a tilted generator for a Markov process on an extended state space.  The size of this state space grows quickly with the model complexity, which makes explicit computations tedious.  We explain here that a milder extension to the state space is already sufficient to obtain the SCGF (at least for non-revisiting cycles).

This (extended) state space contains all the elements of the original space $\Gamma$, along with states corresponding to progressive partial completions $\tilde \CC_2, \ldots, \tilde \CC_{m}$ of any cycle of interest (with length $m$). As an example, we consider the cycles $\CC=\as\bs\cs\as$ and $\CR=\as\cs\bs\as$, in which case the state space is extended by the states $\tilde \CC_2=\as\bs$, $\tilde \CC_3=\as\bs\cs$, and $\tilde \CC^{\rm R}_2=\as\cs$, $\tilde \CC^{\rm R}_3=\as\cs\bs$.  
{A network representation of a Markov process on this extended space is shown in Fig.~\ref{fig:extended_network}.}  Note that this extended state space grows only linearly with the length of the cycle of interest, as opposed to the exponential growth of the corresponding $m$-word space.

\begin{figure}
  \centering
  \includegraphics[width=0.8\linewidth]{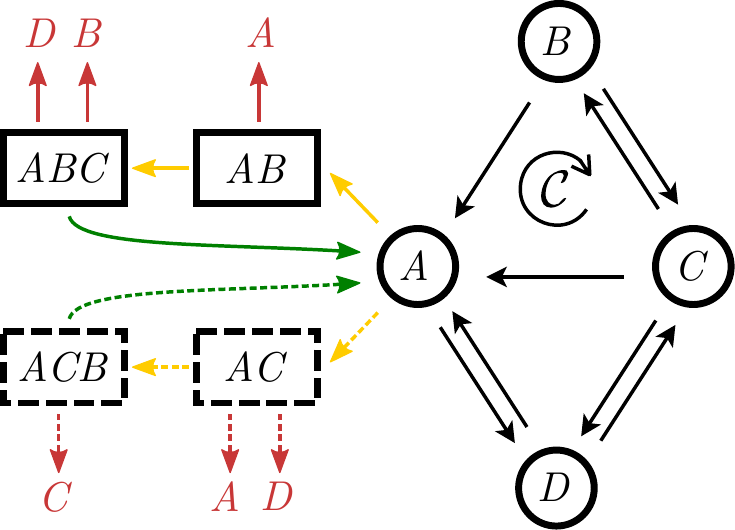}
  \caption{Transitions in the extended state space for counting completions of the cycles $\CC$ and $\CR$ for the network of Fig.~\ref{fig:sketch1}. Transitions within the original state space $\Gamma$ (circled states) are shown in black. Additional states (boxed) are the partial completions of cycles $\CC$ (solid) and $\CR$ (dashed). Yellow arrows indicate attempted completions of cycles. These attempts may fail (via the transitions shown in red) or lead to a successful completion of a cycle (via the transitions shown in green). }
  \label{fig:extended_network}
\end{figure}

We order the three sub-spaces of the extended state space as $(\Gamma, \tilde \CC_2, \ldots, \tilde \CC_{m}, \tilde \CC^{\rm R}_2, \ldots, \tilde \CC^{\rm R}_{m})$ and accordingly construct a rate matrix of the block form
\beq
{\mathbb W}^0=\left(
  \begin{matrix}
    {\mathbb W}_{\Gamma} & {\mathbb W}_{\Gamma\CC} & {\mathbb W}_{\Gamma\CR}\ \\
    {\mathbb W}_{\CC\Gamma} & {\mathbb W}_{\CC} & {\mathbf 0} \\
    {\mathbb W}_{\CR\Gamma} & {\mathbf 0} & {\mathbb W}_{\CR} \\
  \end{matrix}
\right).
\eeq
We use the notation $[{\mathbb W}^0]_{\nu\mu}\equiv W(\mu\to \nu)$ to label off-diagonal elements with column $\mu$ and row $\nu$ of the full matrix ${\mathbb W}^0$ in the extended state space. 

The block ${\mathbb W}_{\Gamma}$ describes transitions within $\Gamma$, that do not mark the start of an attempted cycle, i.e., $W(x\to y)=w(x\to y)$ for all $x,y\in \Gamma$, except for $W(\CC_1\to \CC_2)=0$ and $W(\CR_1\to \CC^{\rm R}_2)=0$. These transitions are marked in black in Fig.~\ref{fig:extended_network}.   [It is understood that $w(\CC\to\CC)=0$ in the formulae of this Section, because transitions only take place between distinct states.]

The blocks ${\mathbb W}_{\CC\Gamma}$ and ${\mathbb W}_{\CR\Gamma}$ have one non-zero entry each, marking the start of an attempted cycle. They have the rates $W(\tilde \CC_1\to \tilde\CC_2)=w(\CC_1\to \CC_2)$ and $W(\CR_1\to \tilde\CC^{\rm R}_2)=w(\CR_1\to \CR_2)$.  Fig.~\ref{fig:extended_network} shows the relevant transitions for example system as yellow arrows leaving $A$.

The blocks ${\mathbb W}_{\CC}$ and ${\mathbb W}_{\CR}$ convey the successful continuation of the attempted cycle. Their non-zero rates are $W(\tilde\CC_i\to \tilde\CC_{i+1})=w(\CC_i\to \CC_{i+1})$ and $W(\tilde\CC^{\rm R}_i\to \tilde\CC^{\rm R}_{i+1})=w(\CR_i\to \CR_{i+1})$ for $2\leq i< m$, corresponding to the other yellow arrows in Fig.~\ref{fig:extended_network}.

The blocks ${\mathbb W}_{\Gamma\CC}$ and $ {\mathbb W}_{\Gamma\CR}$ convey transitions that mark the end of an (attempted) cycle. These are mostly unsuccessful terminations of an attempted cycle (shown in red in Fig.~\ref{fig:extended_network}), except for a single transition for each cycle that closes it correctly (shown in green). The transition rates are $W(\tilde\CC_i\to y)=w(\CC_i\to y)$ for $2\leq i\leq m$ and $y\in\Gamma$, except for $W(\tilde\CC_i\to \CC_{i+1})=0$ when $i<m$; and likewise for $\CR$.

Finally, the diagonal elements of the transition matrix are set to $[{\mathbb W}^0]_{\nu\nu}=-\sum_{\mu\neq\nu}[{\mathbb W}^0]_{\mu\nu}$.
The SCGF is obtained as the largest eigenvalue of the tilted matrix ${\mathbb W}(s,\lambda)  = {\mathbb W}^0 + {\mathbb V}(s,\lambda)$, analogously to Eq.~\eqref{equ:tilted_matrix}, where we count successful transitions from $\tilde\CC_m$ to $\CC_{m+1}$ or from $\tilde\CC^{\rm R}_m$ to $\CR_{m+1}$ by setting
\begin{align}
[{\mathbb V}(s,\lambda)]_{\CC_{m+1}, \tilde\CC_m} & = ({\rm e}^{s+\lambda}-1) [{\mathbb W}^0]_{\CC_{m+1}, \tilde\CC_m} \; ,
\nonumber\\
[{\mathbb V}(s,\lambda)]_{\CR_{m+1}, \tilde\CC^{\rm R}_m} & = ({\rm e}^{s-\lambda}-1) [{\mathbb W}^0]_{\CR_{m+1}, \tilde\CC^{\rm R}_m} \; ,
\end{align}
and all other entries of $[{\mathbb V}(s,\lambda)]_{\nu\mu}$ to zero.

For the particular example of Fig.~\ref{fig:extended_network}, writing $w_{xy}\equiv w(x\to y)$ and $r_x\equiv r(x)$, we obtain the tilted matrix (omitting zero elements):
\begin{align}
  &{\mathbb W}(s,\lambda)=\nonumber\\
  &\left[
  \begin{array}{cccc|cc|cc}
           -r_\as & w_{\bs\as} & w_{\cs\as} & w_{\ds\as} & w_{\bs\as} & w_{\cs\as}{\rm e}^{s+\lambda} & w_{\cs\as} & w_{\bs\as}{\rm e}^{s-\lambda} \\
           &   -r_\bs    & w_{\cs\bs} &        &        & w_{\cs\bs} &       &        \\
           & w_{\bs\cs} &   -r_\cs    & w_{\ds\cs} &        &        &        & w_{\bs\cs} \\
    w_{\as\ds} &        & w_{\cs\ds} &   -r_\ds     &        & w_{\cs\ds} & w_{\cs\ds} &        \\\hline
    w_{\as\bs} &  & & & -r_{\bs} & & & \\
     & & & & w_{\bs\cs} & -r_{\cs} & & \\\hline
    w_{\as\cs} & & & & & &-r_{\cs} & \\
           & & & & & & w_{\cs\bs} & -r_{\bs}
  \end{array}
\right] \; .
\end{align}
If $s,\lambda=0$, this is a stochastic matrix -- its columns sum to zero.
Rows and columns correspond to the extended state space
\beq
(\as,\bs,\cs,\ds,\as\bs,\as\bs\cs,\as\cs,\as\cs\bs).
\eeq
The SCGFs in Fig.~\ref{fig:Psi} were obtained by finding the largest eigenvalue of this matrix (for various $s,\lambda$).

\end{appendix}

\bibliography{cyc}

%merlin.mbs apsrev4-1.bst 2010-07-25 4.21a (PWD, AO, DPC) hacked
%Control: key (0)
%Control: author (8) initials jnrlst
%Control: editor formatted (1) identically to author
%Control: production of article title (-1) disabled
%Control: page (0) single
%Control: year (1) truncated
%Control: production of eprint (0) enabled
\begin{thebibliography}{46}%
\makeatletter
\providecommand \@ifxundefined [1]{%
 \@ifx{#1\undefined}
}%
\providecommand \@ifnum [1]{%
 \ifnum #1\expandafter \@firstoftwo
 \else \expandafter \@secondoftwo
 \fi
}%
\providecommand \@ifx [1]{%
 \ifx #1\expandafter \@firstoftwo
 \else \expandafter \@secondoftwo
 \fi
}%
\providecommand \natexlab [1]{#1}%
\providecommand \enquote  [1]{``#1''}%
\providecommand \bibnamefont  [1]{#1}%
\providecommand \bibfnamefont [1]{#1}%
\providecommand \citenamefont [1]{#1}%
\providecommand \href@noop [0]{\@secondoftwo}%
\providecommand \href [0]{\begingroup \@sanitize@url \@href}%
\providecommand \@href[1]{\@@startlink{#1}\@@href}%
\providecommand \@@href[1]{\endgroup#1\@@endlink}%
\providecommand \@sanitize@url [0]{\catcode `\\12\catcode `\$12\catcode
  `\&12\catcode `\#12\catcode `\^12\catcode `\_12\catcode `\%12\relax}%
\providecommand \@@startlink[1]{}%
\providecommand \@@endlink[0]{}%
\providecommand \url  [0]{\begingroup\@sanitize@url \@url }%
\providecommand \@url [1]{\endgroup\@href {#1}{\urlprefix }}%
\providecommand \urlprefix  [0]{URL }%
\providecommand \Eprint [0]{\href }%
\providecommand \doibase [0]{http://dx.doi.org/}%
\providecommand \selectlanguage [0]{\@gobble}%
\providecommand \bibinfo  [0]{\@secondoftwo}%
\providecommand \bibfield  [0]{\@secondoftwo}%
\providecommand \translation [1]{[#1]}%
\providecommand \BibitemOpen [0]{}%
\providecommand \bibitemStop [0]{}%
\providecommand \bibitemNoStop [0]{.\EOS\space}%
\providecommand \EOS [0]{\spacefactor3000\relax}%
\providecommand \BibitemShut  [1]{\csname bibitem#1\endcsname}%
\let\auto@bib@innerbib\@empty
%</preamble>
\bibitem [{\citenamefont {Gallavotti}\ and\ \citenamefont
  {Cohen}(1995)}]{Gallavotti1995}%
  \BibitemOpen
  \bibfield  {author} {\bibinfo {author} {\bibfnamefont {G.}~\bibnamefont
  {Gallavotti}}\ and\ \bibinfo {author} {\bibfnamefont {E.~G.~D.}\ \bibnamefont
  {Cohen}},\ }\href@noop {} {\bibfield  {journal} {\bibinfo  {journal} {J.
  Stat. Phys.}\ }\textbf {\bibinfo {volume} {80}},\ \bibinfo {pages} {931}
  (\bibinfo {year} {1995})}\BibitemShut {NoStop}%
\bibitem [{\citenamefont {Jarzynski}(1997)}]{Jarzynski1997}%
  \BibitemOpen
  \bibfield  {author} {\bibinfo {author} {\bibfnamefont {C.}~\bibnamefont
  {Jarzynski}},\ }\href@noop {} {\bibfield  {journal} {\bibinfo  {journal}
  {Phys. Rev. Lett.}\ }\textbf {\bibinfo {volume} {78}},\ \bibinfo {pages}
  {2690} (\bibinfo {year} {1997})}\BibitemShut {NoStop}%
\bibitem [{\citenamefont {Crooks}(2000)}]{Crooks2000}%
  \BibitemOpen
  \bibfield  {author} {\bibinfo {author} {\bibfnamefont {G.~E.}\ \bibnamefont
  {Crooks}},\ }\href@noop {} {\bibfield  {journal} {\bibinfo  {journal} {Phys.
  Rev. E}\ }\textbf {\bibinfo {volume} {61}},\ \bibinfo {pages} {2361}
  (\bibinfo {year} {2000})}\BibitemShut {NoStop}%
\bibitem [{\citenamefont {Andrieux}\ and\ \citenamefont {Gaspard}(2007)}]{AG}%
  \BibitemOpen
  \bibfield  {author} {\bibinfo {author} {\bibfnamefont {D.}~\bibnamefont
  {Andrieux}}\ and\ \bibinfo {author} {\bibfnamefont {P.}~\bibnamefont
  {Gaspard}},\ }\href@noop {} {\bibfield  {journal} {\bibinfo  {journal} {J.
  Stat. Phys.}\ }\textbf {\bibinfo {volume} {127}},\ \bibinfo {pages} {107}
  (\bibinfo {year} {2007})}\BibitemShut {NoStop}%
\bibitem [{\citenamefont {Seifert}(2012)}]{Seifert2012}%
  \BibitemOpen
  \bibfield  {author} {\bibinfo {author} {\bibfnamefont {U.}~\bibnamefont
  {Seifert}},\ }\href@noop {} {\bibfield  {journal} {\bibinfo  {journal} {Rep.
  Prog. Phys.}\ }\textbf {\bibinfo {volume} {75}},\ \bibinfo {pages} {126001}
  (\bibinfo {year} {2012})}\BibitemShut {NoStop}%
\bibitem [{\citenamefont {Barato}\ and\ \citenamefont
  {Seifert}(2015{\natexlab{a}})}]{bara15}%
  \BibitemOpen
  \bibfield  {author} {\bibinfo {author} {\bibfnamefont {A.~C.}\ \bibnamefont
  {Barato}}\ and\ \bibinfo {author} {\bibfnamefont {U.}~\bibnamefont
  {Seifert}},\ }\href {\doibase 10.1103/PhysRevLett.114.158101} {\bibfield
  {journal} {\bibinfo  {journal} {Phys. Rev. Lett.}\ }\textbf {\bibinfo
  {volume} {114}},\ \bibinfo {pages} {158101} (\bibinfo {year}
  {2015}{\natexlab{a}})}\BibitemShut {NoStop}%
\bibitem [{\citenamefont {Gingrich}\ \emph {et~al.}(2016)\citenamefont
  {Gingrich}, \citenamefont {Horowitz}, \citenamefont {Perunov},\ and\
  \citenamefont {England}}]{ging16}%
  \BibitemOpen
  \bibfield  {author} {\bibinfo {author} {\bibfnamefont {T.~R.}\ \bibnamefont
  {Gingrich}}, \bibinfo {author} {\bibfnamefont {J.~M.}\ \bibnamefont
  {Horowitz}}, \bibinfo {author} {\bibfnamefont {N.}~\bibnamefont {Perunov}}, \
  and\ \bibinfo {author} {\bibfnamefont {J.~L.}\ \bibnamefont {England}},\
  }\href {\doibase 10.1103/PhysRevLett.116.120601} {\bibfield  {journal}
  {\bibinfo  {journal} {Phys. Rev. Lett.}\ }\textbf {\bibinfo {volume} {116}},\
  \bibinfo {pages} {120601} (\bibinfo {year} {2016})}\BibitemShut {NoStop}%
\bibitem [{\citenamefont {Pietzonka}\ \emph {et~al.}(2016)\citenamefont
  {Pietzonka}, \citenamefont {Barato},\ and\ \citenamefont {Seifert}}]{piet15}%
  \BibitemOpen
  \bibfield  {author} {\bibinfo {author} {\bibfnamefont {P.}~\bibnamefont
  {Pietzonka}}, \bibinfo {author} {\bibfnamefont {A.~C.}\ \bibnamefont
  {Barato}}, \ and\ \bibinfo {author} {\bibfnamefont {U.}~\bibnamefont
  {Seifert}},\ }\href {\doibase 10.1103/PhysRevE.93.052145} {\bibfield
  {journal} {\bibinfo  {journal} {Phys. Rev. E}\ }\textbf {\bibinfo {volume}
  {93}},\ \bibinfo {pages} {052145} (\bibinfo {year} {2016})}\BibitemShut
  {NoStop}%
\bibitem [{\citenamefont {Horowitz}\ and\ \citenamefont
  {Gingrich}(2019)}]{horo19}%
  \BibitemOpen
  \bibfield  {author} {\bibinfo {author} {\bibfnamefont {J.~M.}\ \bibnamefont
  {Horowitz}}\ and\ \bibinfo {author} {\bibfnamefont {T.~R.}\ \bibnamefont
  {Gingrich}},\ }\href@noop {} {\bibfield  {journal} {\bibinfo  {journal} {Nat.
  Phys.}\ }\textbf {\bibinfo {volume} {16}},\ \bibinfo {pages} {15} (\bibinfo
  {year} {2019})}\BibitemShut {NoStop}%
\bibitem [{\citenamefont {Dechant}\ and\ \citenamefont
  {Sasa}(2020)}]{Dechant2020}%
  \BibitemOpen
  \bibfield  {author} {\bibinfo {author} {\bibfnamefont {A.}~\bibnamefont
  {Dechant}}\ and\ \bibinfo {author} {\bibfnamefont {S.-i.}\ \bibnamefont
  {Sasa}},\ }\href@noop {} {\bibfield  {journal} {\bibinfo  {journal} {Proc.
  Nat. Acad. Sci. USA}\ }\textbf {\bibinfo {volume} {117}},\ \bibinfo {pages}
  {6430} (\bibinfo {year} {2020})}\BibitemShut {NoStop}%
\bibitem [{\citenamefont {Liphardt}\ \emph {et~al.}(2002)\citenamefont
  {Liphardt}, \citenamefont {Dumont}, \citenamefont {Smith}, \citenamefont
  {Tinoco},\ and\ \citenamefont {Bustamante}}]{Liphardt2002}%
  \BibitemOpen
  \bibfield  {author} {\bibinfo {author} {\bibfnamefont {J.}~\bibnamefont
  {Liphardt}}, \bibinfo {author} {\bibfnamefont {S.}~\bibnamefont {Dumont}},
  \bibinfo {author} {\bibfnamefont {S.~B.}\ \bibnamefont {Smith}}, \bibinfo
  {author} {\bibfnamefont {I.}~\bibnamefont {Tinoco}}, \ and\ \bibinfo {author}
  {\bibfnamefont {C.}~\bibnamefont {Bustamante}},\ }\href@noop {} {\bibfield
  {journal} {\bibinfo  {journal} {Science}\ }\textbf {\bibinfo {volume}
  {296}},\ \bibinfo {pages} {1832} (\bibinfo {year} {2002})}\BibitemShut
  {NoStop}%
\bibitem [{\citenamefont {Ciliberto}(2017)}]{cili17}%
  \BibitemOpen
  \bibfield  {author} {\bibinfo {author} {\bibfnamefont {S.}~\bibnamefont
  {Ciliberto}},\ }\href {\doibase 10.1103/PhysRevX.7.021051} {\bibfield
  {journal} {\bibinfo  {journal} {Phys. Rev. X}\ }\textbf {\bibinfo {volume}
  {7}},\ \bibinfo {pages} {021051} (\bibinfo {year} {2017})}\BibitemShut
  {NoStop}%
\bibitem [{\citenamefont {Barato}\ and\ \citenamefont
  {Seifert}(2015{\natexlab{b}})}]{bara15a}%
  \BibitemOpen
  \bibfield  {author} {\bibinfo {author} {\bibfnamefont {A.~C.}\ \bibnamefont
  {Barato}}\ and\ \bibinfo {author} {\bibfnamefont {U.}~\bibnamefont
  {Seifert}},\ }\href {\doibase 10.1021/acs.jpcb.5b01918} {\bibfield  {journal}
  {\bibinfo  {journal} {J. Phys. Chem. B}\ }\textbf {\bibinfo {volume} {119}},\
  \bibinfo {pages} {6555} (\bibinfo {year} {2015}{\natexlab{b}})}\BibitemShut
  {NoStop}%
\bibitem [{\citenamefont {Li}\ \emph {et~al.}(2019)\citenamefont {Li},
  \citenamefont {Horowitz}, \citenamefont {Gingrich},\ and\ \citenamefont
  {Fakhri}}]{li19}%
  \BibitemOpen
  \bibfield  {author} {\bibinfo {author} {\bibfnamefont {J.}~\bibnamefont
  {Li}}, \bibinfo {author} {\bibfnamefont {J.~M.}\ \bibnamefont {Horowitz}},
  \bibinfo {author} {\bibfnamefont {T.~R.}\ \bibnamefont {Gingrich}}, \ and\
  \bibinfo {author} {\bibfnamefont {N.}~\bibnamefont {Fakhri}},\ }\href
  {\doibase 10.1038/s41467-019-09631-x} {\bibfield  {journal} {\bibinfo
  {journal} {Nat. Commun.}\ }\textbf {\bibinfo {volume} {10}},\ \bibinfo
  {pages} {1666} (\bibinfo {year} {2019})}\BibitemShut {NoStop}%
\bibitem [{\citenamefont {Manikandan}\ \emph {et~al.}(2020)\citenamefont
  {Manikandan}, \citenamefont {Gupta},\ and\ \citenamefont
  {Krishnamurthy}}]{mani20}%
  \BibitemOpen
  \bibfield  {author} {\bibinfo {author} {\bibfnamefont {S.~K.}\ \bibnamefont
  {Manikandan}}, \bibinfo {author} {\bibfnamefont {D.}~\bibnamefont {Gupta}}, \
  and\ \bibinfo {author} {\bibfnamefont {S.}~\bibnamefont {Krishnamurthy}},\
  }\href {\doibase 10.1103/PhysRevLett.124.120603} {\bibfield  {journal}
  {\bibinfo  {journal} {Phys. Rev. Lett.}\ }\textbf {\bibinfo {volume} {124}},\
  \bibinfo {pages} {120603} (\bibinfo {year} {2020})}\BibitemShut {NoStop}%
\bibitem [{\citenamefont {Biddle}\ and\ \citenamefont
  {Gunawardena}(2020)}]{BG}%
  \BibitemOpen
  \bibfield  {author} {\bibinfo {author} {\bibfnamefont {J.~W.}\ \bibnamefont
  {Biddle}}\ and\ \bibinfo {author} {\bibfnamefont {J.}~\bibnamefont
  {Gunawardena}},\ }\href {\doibase 10.1103/PhysRevE.101.062125} {\bibfield
  {journal} {\bibinfo  {journal} {Phys. Rev. E}\ }\textbf {\bibinfo {volume}
  {101}},\ \bibinfo {pages} {062125} (\bibinfo {year} {2020})}\BibitemShut
  {NoStop}%
\bibitem [{\citenamefont {Jiang}\ \emph {et~al.}(2004)\citenamefont {Jiang},
  \citenamefont {Qian},\ and\ \citenamefont {Qian}}]{qian-book}%
  \BibitemOpen
  \bibfield  {author} {\bibinfo {author} {\bibfnamefont {D.-Q.}\ \bibnamefont
  {Jiang}}, \bibinfo {author} {\bibfnamefont {M.}~\bibnamefont {Qian}}, \ and\
  \bibinfo {author} {\bibfnamefont {M.-P.}\ \bibnamefont {Qian}},\ }\href@noop
  {} {\emph {\bibinfo {title} {Mathematical Theory of Nonequilibrium Steady
  States}}}\ (\bibinfo  {publisher} {Springer},\ \bibinfo {address}
  {Berlin/Heidelberg},\ \bibinfo {year} {2004})\BibitemShut {NoStop}%
\bibitem [{\citenamefont {Kalpazidou}(1995)}]{kalpazidou-book}%
  \BibitemOpen
  \bibfield  {author} {\bibinfo {author} {\bibfnamefont {S.~L.}\ \bibnamefont
  {Kalpazidou}},\ }\href@noop {} {\emph {\bibinfo {title} {Cycle
  Representations of Markov Processes}}}\ (\bibinfo  {publisher} {Springer},\
  \bibinfo {year} {1995})\BibitemShut {NoStop}%
\bibitem [{\citenamefont {Jia}\ \emph {et~al.}(2016)\citenamefont {Jia},
  \citenamefont {Jiang},\ and\ \citenamefont {Qian}}]{Jia2016}%
  \BibitemOpen
  \bibfield  {author} {\bibinfo {author} {\bibfnamefont {C.}~\bibnamefont
  {Jia}}, \bibinfo {author} {\bibfnamefont {D.-Q.}\ \bibnamefont {Jiang}}, \
  and\ \bibinfo {author} {\bibfnamefont {M.-P.}\ \bibnamefont {Qian}},\
  }\href@noop {} {\bibfield  {journal} {\bibinfo  {journal} {Ann. Appl. Prob.}\
  }\textbf {\bibinfo {volume} {26}},\ \bibinfo {pages} {2454} (\bibinfo {year}
  {2016})}\BibitemShut {NoStop}%
\bibitem [{\citenamefont {Polettini}\ \emph {et~al.}()\citenamefont
  {Polettini}, \citenamefont {Falasco},\ and\ \citenamefont
  {Esposito}}]{polettini2021}%
  \BibitemOpen
  \bibfield  {author} {\bibinfo {author} {\bibfnamefont {M.}~\bibnamefont
  {Polettini}}, \bibinfo {author} {\bibfnamefont {G.}~\bibnamefont {Falasco}},
  \ and\ \bibinfo {author} {\bibfnamefont {M.}~\bibnamefont {Esposito}},\
  }\href@noop {} {}\bibinfo {howpublished} {arxiv:2106.00425}\BibitemShut
  {NoStop}%
\bibitem [{\citenamefont {Schnakenberg}(1976)}]{schn76}%
  \BibitemOpen
  \bibfield  {author} {\bibinfo {author} {\bibfnamefont {J.}~\bibnamefont
  {Schnakenberg}},\ }\href {\doibase 10.1103/RevModPhys.48.571} {\bibfield
  {journal} {\bibinfo  {journal} {Rev. Mod. Phys.}\ }\textbf {\bibinfo {volume}
  {48}},\ \bibinfo {pages} {571} (\bibinfo {year} {1976})}\BibitemShut
  {NoStop}%
\bibitem [{\citenamefont {Altaner}\ \emph {et~al.}(2012)\citenamefont
  {Altaner}, \citenamefont {Grosskinsky}, \citenamefont {Herminghaus},
  \citenamefont {Katth\"an}, \citenamefont {Timme},\ and\ \citenamefont
  {Vollmer}}]{Altaner2012}%
  \BibitemOpen
  \bibfield  {author} {\bibinfo {author} {\bibfnamefont {B.}~\bibnamefont
  {Altaner}}, \bibinfo {author} {\bibfnamefont {S.}~\bibnamefont
  {Grosskinsky}}, \bibinfo {author} {\bibfnamefont {S.}~\bibnamefont
  {Herminghaus}}, \bibinfo {author} {\bibfnamefont {L.}~\bibnamefont
  {Katth\"an}}, \bibinfo {author} {\bibfnamefont {M.}~\bibnamefont {Timme}}, \
  and\ \bibinfo {author} {\bibfnamefont {J.}~\bibnamefont {Vollmer}},\
  }\href@noop {} {\bibfield  {journal} {\bibinfo  {journal} {Phys. Rev. E}\
  }\textbf {\bibinfo {volume} {85}},\ \bibinfo {pages} {041133} (\bibinfo
  {year} {2012})}\BibitemShut {NoStop}%
\bibitem [{\citenamefont {Bertini}\ \emph
  {et~al.}(2015{\natexlab{a}})\citenamefont {Bertini}, \citenamefont
  {Faggionato},\ and\ \citenamefont {Gabrielli}}]{Bertini2015-spa}%
  \BibitemOpen
  \bibfield  {author} {\bibinfo {author} {\bibfnamefont {L.}~\bibnamefont
  {Bertini}}, \bibinfo {author} {\bibfnamefont {A.}~\bibnamefont {Faggionato}},
  \ and\ \bibinfo {author} {\bibfnamefont {D.}~\bibnamefont {Gabrielli}},\
  }\href@noop {} {\bibfield  {journal} {\bibinfo  {journal} {Stochastic
  Process. Appl.}\ }\textbf {\bibinfo {volume} {125}},\ \bibinfo {pages} {2786}
  (\bibinfo {year} {2015}{\natexlab{a}})}\BibitemShut {NoStop}%
\bibitem [{\citenamefont {Schbath}(1997)}]{Schbath1997}%
  \BibitemOpen
  \bibfield  {author} {\bibinfo {author} {\bibfnamefont {S.}~\bibnamefont
  {Schbath}},\ }\href {\doibase 10.1051/ps:1997100} {\bibfield  {journal}
  {\bibinfo  {journal} {ESAIM: Probability and Statistics}\ }\textbf {\bibinfo
  {volume} {1}},\ \bibinfo {pages} {1} (\bibinfo {year} {1997})}\BibitemShut
  {NoStop}%
\bibitem [{\citenamefont {Robin}\ and\ \citenamefont
  {Daudin}(1999)}]{robin_daudin_1999}%
  \BibitemOpen
  \bibfield  {author} {\bibinfo {author} {\bibfnamefont {S.}~\bibnamefont
  {Robin}}\ and\ \bibinfo {author} {\bibfnamefont {J.~J.}\ \bibnamefont
  {Daudin}},\ }\href {\doibase 10.1239/jap/1032374240} {\bibfield  {journal}
  {\bibinfo  {journal} {J. Appl. Prob.}\ }\textbf {\bibinfo {volume} {36}},\
  \bibinfo {pages} {179} (\bibinfo {year} {1999})}\BibitemShut {NoStop}%
\bibitem [{\citenamefont {Lothaire}(2005)}]{Lothaire2005}%
  \BibitemOpen
  \bibfield  {author} {\bibinfo {author} {\bibfnamefont {M.}~\bibnamefont
  {Lothaire}},\ }\enquote {\bibinfo {title} {Statistics on words with
  applications to biological sequences},}\ in\ \href {\doibase DOI:
  10.1017/CBO9781107341005.007} {\emph {\bibinfo {booktitle} {Applied
  Combinatorics on Words}}},\ \bibinfo {series and number} {Encyclopedia of
  Mathematics and its Applications}\ (\bibinfo  {publisher} {Cambridge
  University Press},\ \bibinfo {address} {Cambridge},\ \bibinfo {year} {2005})\
  pp.\ \bibinfo {pages} {268--352}\BibitemShut {NoStop}%
\bibitem [{\citenamefont {Roquain}\ and\ \citenamefont
  {Schbath}(2007)}]{Roquain2007}%
  \BibitemOpen
  \bibfield  {author} {\bibinfo {author} {\bibfnamefont {E.}~\bibnamefont
  {Roquain}}\ and\ \bibinfo {author} {\bibfnamefont {S.}~\bibnamefont
  {Schbath}},\ }\href {\doibase DOI: 10.1239/aap/1175266472} {\bibfield
  {journal} {\bibinfo  {journal} {Adv. Appl. Prob.}\ }\textbf {\bibinfo
  {volume} {39}},\ \bibinfo {pages} {128} (\bibinfo {year} {2007})}\BibitemShut
  {NoStop}%
\bibitem [{\citenamefont {Rold{\'a}n}\ and\ \citenamefont
  {Vivo}(2019)}]{Roldan2019}%
  \BibitemOpen
  \bibfield  {author} {\bibinfo {author} {\bibfnamefont {{\'E}.}~\bibnamefont
  {Rold{\'a}n}}\ and\ \bibinfo {author} {\bibfnamefont {P.}~\bibnamefont
  {Vivo}},\ }\href@noop {} {\bibfield  {journal} {\bibinfo  {journal} {Phys.
  Rev. E}\ }\textbf {\bibinfo {volume} {100}},\ \bibinfo {pages} {042108}
  (\bibinfo {year} {2019})}\BibitemShut {NoStop}%
\bibitem [{\citenamefont {Touchette}(2009)}]{Touchette2009}%
  \BibitemOpen
  \bibfield  {author} {\bibinfo {author} {\bibfnamefont {H.}~\bibnamefont
  {Touchette}},\ }\href@noop {} {\bibfield  {journal} {\bibinfo  {journal}
  {Phys. Rep.}\ }\textbf {\bibinfo {volume} {478}},\ \bibinfo {pages} {1}
  (\bibinfo {year} {2009})}\BibitemShut {NoStop}%
\bibitem [{\citenamefont {Ch\'{e}trite}\ and\ \citenamefont
  {Touchette}(2015)}]{Chetrite2015}%
  \BibitemOpen
  \bibfield  {author} {\bibinfo {author} {\bibfnamefont {R.}~\bibnamefont
  {Ch\'{e}trite}}\ and\ \bibinfo {author} {\bibfnamefont {H.}~\bibnamefont
  {Touchette}},\ }\href@noop {} {\bibfield  {journal} {\bibinfo  {journal}
  {Ann. Henri Poincar\'{e}}\ }\textbf {\bibinfo {volume} {16}},\ \bibinfo
  {pages} {2005} (\bibinfo {year} {2015})}\BibitemShut {NoStop}%
\bibitem [{\citenamefont {Jack}(2020)}]{Jack2020}%
  \BibitemOpen
  \bibfield  {author} {\bibinfo {author} {\bibfnamefont {R.~L.}\ \bibnamefont
  {Jack}},\ }\href {\doibase 10.1140/epjb/e2020-100605-3} {\bibfield  {journal}
  {\bibinfo  {journal} {Eur. Phys. J. B}\ }\textbf {\bibinfo {volume} {93}},\
  \bibinfo {pages} {74} (\bibinfo {year} {2020})}\BibitemShut {NoStop}%
\bibitem [{\citenamefont {Maes}(2020)}]{Maes2020}%
  \BibitemOpen
  \bibfield  {author} {\bibinfo {author} {\bibfnamefont {C.}~\bibnamefont
  {Maes}},\ }\href@noop {} {\bibfield  {journal} {\bibinfo  {journal} {Physics
  Reports}\ }\textbf {\bibinfo {volume} {850}},\ \bibinfo {pages} {1} (\bibinfo
  {year} {2020})}\BibitemShut {NoStop}%
\bibitem [{\citenamefont {Ge}(2008)}]{Ge2008}%
  \BibitemOpen
  \bibfield  {author} {\bibinfo {author} {\bibfnamefont {H.}~\bibnamefont
  {Ge}},\ }\href@noop {} {\bibfield  {journal} {\bibinfo  {journal} {J. Phys.
  Chem. B}\ }\textbf {\bibinfo {volume} {112}},\ \bibinfo {pages} {61}
  (\bibinfo {year} {2008})}\BibitemShut {NoStop}%
\bibitem [{\citenamefont {Ge}(2012)}]{Ge2012}%
  \BibitemOpen
  \bibfield  {author} {\bibinfo {author} {\bibfnamefont {H.}~\bibnamefont
  {Ge}},\ }\href@noop {} {\bibfield  {journal} {\bibinfo  {journal} {J. Phys.
  A.}\ }\textbf {\bibinfo {volume} {45}},\ \bibinfo {pages} {215002} (\bibinfo
  {year} {2012})}\BibitemShut {NoStop}%
\bibitem [{\citenamefont {Maes}\ and\ \citenamefont
  {Neto{\v{c}}n{\'{y}}}(2008)}]{maes08}%
  \BibitemOpen
  \bibfield  {author} {\bibinfo {author} {\bibfnamefont {C.}~\bibnamefont
  {Maes}}\ and\ \bibinfo {author} {\bibfnamefont {K.}~\bibnamefont
  {Neto{\v{c}}n{\'{y}}}},\ }\href {\doibase 10.1209/0295-5075/82/30003}
  {\bibfield  {journal} {\bibinfo  {journal} {{EPL}}\ }\textbf {\bibinfo
  {volume} {82}},\ \bibinfo {pages} {30003} (\bibinfo {year}
  {2008})}\BibitemShut {NoStop}%
\bibitem [{\citenamefont {Mehl}\ \emph {et~al.}(2012)\citenamefont {Mehl},
  \citenamefont {Lander}, \citenamefont {Bechinger}, \citenamefont {Blickle},\
  and\ \citenamefont {Seifert}}]{mehl12}%
  \BibitemOpen
  \bibfield  {author} {\bibinfo {author} {\bibfnamefont {J.}~\bibnamefont
  {Mehl}}, \bibinfo {author} {\bibfnamefont {B.}~\bibnamefont {Lander}},
  \bibinfo {author} {\bibfnamefont {C.}~\bibnamefont {Bechinger}}, \bibinfo
  {author} {\bibfnamefont {V.}~\bibnamefont {Blickle}}, \ and\ \bibinfo
  {author} {\bibfnamefont {U.}~\bibnamefont {Seifert}},\ }\href {\doibase
  10.1103/PhysRevLett.108.220601} {\bibfield  {journal} {\bibinfo  {journal}
  {Phys. Rev. Lett.}\ }\textbf {\bibinfo {volume} {108}},\ \bibinfo {pages}
  {220601} (\bibinfo {year} {2012})}\BibitemShut {NoStop}%
\bibitem [{\citenamefont {Polettini}\ and\ \citenamefont
  {Esposito}(2017)}]{pole17}%
  \BibitemOpen
  \bibfield  {author} {\bibinfo {author} {\bibfnamefont {M.}~\bibnamefont
  {Polettini}}\ and\ \bibinfo {author} {\bibfnamefont {M.}~\bibnamefont
  {Esposito}},\ }\href {\doibase 10.1103/PhysRevLett.119.240601} {\bibfield
  {journal} {\bibinfo  {journal} {Phys. Rev. Lett.}\ }\textbf {\bibinfo
  {volume} {119}},\ \bibinfo {pages} {240601} (\bibinfo {year}
  {2017})}\BibitemShut {NoStop}%
\bibitem [{\citenamefont {Uhl}\ \emph {et~al.}(2018)\citenamefont {Uhl},
  \citenamefont {Pietzonka},\ and\ \citenamefont {Seifert}}]{uhl18}%
  \BibitemOpen
  \bibfield  {author} {\bibinfo {author} {\bibfnamefont {M.}~\bibnamefont
  {Uhl}}, \bibinfo {author} {\bibfnamefont {P.}~\bibnamefont {Pietzonka}}, \
  and\ \bibinfo {author} {\bibfnamefont {U.}~\bibnamefont {Seifert}},\ }\href
  {\doibase 10.1088/1742-5468/aaa78b} {\bibfield  {journal} {\bibinfo
  {journal} {J. Stat. Mech. Theory Exp.}\ ,\ \bibinfo {pages} {023203}}
  (\bibinfo {year} {2018})}\BibitemShut {NoStop}%
\bibitem [{\citenamefont {Kahlen}\ and\ \citenamefont {Ehrich}(2018)}]{kahl18}%
  \BibitemOpen
  \bibfield  {author} {\bibinfo {author} {\bibfnamefont {M.}~\bibnamefont
  {Kahlen}}\ and\ \bibinfo {author} {\bibfnamefont {J.}~\bibnamefont
  {Ehrich}},\ }\href {\doibase 10.1088/1742-5468/aac2fd} {\bibfield  {journal}
  {\bibinfo  {journal} {J. Stat. Mech. Theory Exp.}\ }\textbf {\bibinfo
  {volume} {2018}},\ \bibinfo {pages} {063204} (\bibinfo {year}
  {2018})}\BibitemShut {NoStop}%
\bibitem [{\citenamefont {Hayashi}\ \emph {et~al.}(2010)\citenamefont
  {Hayashi}, \citenamefont {Ueno}, \citenamefont {Iino},\ and\ \citenamefont
  {Noji}}]{haya10}%
  \BibitemOpen
  \bibfield  {author} {\bibinfo {author} {\bibfnamefont {K.}~\bibnamefont
  {Hayashi}}, \bibinfo {author} {\bibfnamefont {H.}~\bibnamefont {Ueno}},
  \bibinfo {author} {\bibfnamefont {R.}~\bibnamefont {Iino}}, \ and\ \bibinfo
  {author} {\bibfnamefont {H.}~\bibnamefont {Noji}},\ }\href {\doibase
  10.1103/PhysRevLett.104.218103} {\bibfield  {journal} {\bibinfo  {journal}
  {Phys. Rev. Lett.}\ }\textbf {\bibinfo {volume} {104}},\ \bibinfo {pages}
  {218103} (\bibinfo {year} {2010})}\BibitemShut {NoStop}%
\bibitem [{\citenamefont {Alemany}\ \emph {et~al.}(2015)\citenamefont
  {Alemany}, \citenamefont {Ribezzi-Crivellari},\ and\ \citenamefont
  {Ritort}}]{alem15}%
  \BibitemOpen
  \bibfield  {author} {\bibinfo {author} {\bibfnamefont {A.}~\bibnamefont
  {Alemany}}, \bibinfo {author} {\bibfnamefont {M.}~\bibnamefont
  {Ribezzi-Crivellari}}, \ and\ \bibinfo {author} {\bibfnamefont
  {F.}~\bibnamefont {Ritort}},\ }\href {\doibase 10.1088/1367-2630/17/7/075009}
  {\bibfield  {journal} {\bibinfo  {journal} {New J. Phys.}\ }\textbf {\bibinfo
  {volume} {17}},\ \bibinfo {pages} {075009} (\bibinfo {year}
  {2015})}\BibitemShut {NoStop}%
\bibitem [{\citenamefont {Lecomte}\ \emph {et~al.}(2007)\citenamefont
  {Lecomte}, \citenamefont {Appert-Rolland},\ and\ \citenamefont {van
  Wijland}}]{Lecomte2007}%
  \BibitemOpen
  \bibfield  {author} {\bibinfo {author} {\bibfnamefont {V.}~\bibnamefont
  {Lecomte}}, \bibinfo {author} {\bibfnamefont {C.}~\bibnamefont
  {Appert-Rolland}}, \ and\ \bibinfo {author} {\bibfnamefont {F.}~\bibnamefont
  {van Wijland}},\ }\href@noop {} {\bibfield  {journal} {\bibinfo  {journal}
  {J. Stat. Phys.}\ }\textbf {\bibinfo {volume} {127}},\ \bibinfo {pages} {51}
  (\bibinfo {year} {2007})}\BibitemShut {NoStop}%
\bibitem [{\citenamefont {Bertini}\ \emph
  {et~al.}(2015{\natexlab{b}})\citenamefont {Bertini}, \citenamefont
  {Faggionato},\ and\ \citenamefont {Gabrielli}}]{Bertini2015b}%
  \BibitemOpen
  \bibfield  {author} {\bibinfo {author} {\bibfnamefont {L.}~\bibnamefont
  {Bertini}}, \bibinfo {author} {\bibfnamefont {A.}~\bibnamefont {Faggionato}},
  \ and\ \bibinfo {author} {\bibfnamefont {D.}~\bibnamefont {Gabrielli}},\
  }\href {\doibase 10.1214/14-AIHP601} {\bibfield  {journal} {\bibinfo
  {journal} {Ann. Inst. Henri Poincar\'e Probab. Stat.}\ }\textbf {\bibinfo
  {volume} {51}},\ \bibinfo {pages} {867} (\bibinfo {year}
  {2015}{\natexlab{b}})}\BibitemShut {NoStop}%
\bibitem [{\citenamefont {van Horssen}\ and\ \citenamefont {Gu{\c t}{\u
  a}}(2015)}]{Horssen2015}%
  \BibitemOpen
  \bibfield  {author} {\bibinfo {author} {\bibfnamefont {M.}~\bibnamefont {van
  Horssen}}\ and\ \bibinfo {author} {\bibfnamefont {M.}~\bibnamefont {Gu{\c
  t}{\u a}}},\ }\href@noop {} {\bibfield  {journal} {\bibinfo  {journal} {J.
  Math. Phys.}\ }\textbf {\bibinfo {volume} {56}},\ \bibinfo {pages} {022109}
  (\bibinfo {year} {2015})}\BibitemShut {NoStop}%
\bibitem [{\citenamefont {Garrahan}\ \emph {et~al.}(2009)\citenamefont
  {Garrahan}, \citenamefont {Jack}, \citenamefont {Lecomte}, \citenamefont
  {Pitard}, \citenamefont {van Duijvendijk},\ and\ \citenamefont {van
  Wijland}}]{Garrahan2009}%
  \BibitemOpen
  \bibfield  {author} {\bibinfo {author} {\bibfnamefont {J.~P.}\ \bibnamefont
  {Garrahan}}, \bibinfo {author} {\bibfnamefont {R.~L.}\ \bibnamefont {Jack}},
  \bibinfo {author} {\bibfnamefont {V.}~\bibnamefont {Lecomte}}, \bibinfo
  {author} {\bibfnamefont {E.}~\bibnamefont {Pitard}}, \bibinfo {author}
  {\bibfnamefont {K.}~\bibnamefont {van Duijvendijk}}, \ and\ \bibinfo {author}
  {\bibfnamefont {F.}~\bibnamefont {van Wijland}},\ }\href@noop {} {\bibfield
  {journal} {\bibinfo  {journal} {J. Phys. A}\ }\textbf {\bibinfo {volume}
  {42}},\ \bibinfo {pages} {075007} (\bibinfo {year} {2009})}\BibitemShut
  {NoStop}%
\bibitem [{\citenamefont {Ch{\'e}trite}\ \emph {et~al.}()\citenamefont
  {Ch{\'e}trite}, \citenamefont {Faggionato},\ and\ \citenamefont
  {Gabrielli}}]{Chetrite-unpub}%
  \BibitemOpen
  \bibfield  {author} {\bibinfo {author} {\bibfnamefont {R.}~\bibnamefont
  {Ch{\'e}trite}}, \bibinfo {author} {\bibfnamefont {A.}~\bibnamefont
  {Faggionato}}, \ and\ \bibinfo {author} {\bibfnamefont {D.}~\bibnamefont
  {Gabrielli}},\ }\href@noop {} {}\bibinfo {howpublished}
  {(unpublished)}\BibitemShut {NoStop}%
\end{thebibliography}%

\end{document}